\documentclass{aa}

\usepackage{graphicx}
\usepackage{txfonts}
\usepackage{lipsum}
\usepackage{subcaption}
\usepackage{lscape}
\usepackage{placeins}

\usepackage{natbib}
\bibpunct{(}{)}{;}{a}{}{,}

\def\int{{\rm INTEGRAL}}

\newcommand{\xmm}{XMM--Newton}
\newcommand{\sw}{Swift}

\def\approxgt{\mathrel{\hbox{\rlap{\lower.55ex \hbox {$\sim$}}
        \kern-.3em \raise.4ex \hbox{$>$}}}}
\def\approxlt{\mathrel{\hbox{\rlap{\lower.55ex \hbox {$\sim$}}
        \kern-.3em \raise.4ex \hbox{$<$}}}}

\def\mdot {\dot M}

\def\ltsima{$\; \buildrel < \over \sim \;$}
\def\lsim{\lower.5ex\hbox{\ltsima}}
\def\gtsima{$\; \buildrel > \over \sim \;$}
\def\gsim{\lower.5ex\hbox{\gtsima}}

\def\hcm {\hbox {\ifmmode $ atom cm$^{-2}\else atom cm$^{-2}$\fi}}

\def \apj {ApJ}

\def \apjl {ApJL}
\def \apjs {ApJS}
\def \aap {A\&A}

\def \apss {A\&Sp Sc.}

\def \mnras {MNRAS}

\def \araa {ARA\&A}

\def \ssr {Space Science Reviews}

\def \nar {New Astronomy Reviews}

\newcommand{\be}{\begin{equation}}

\newcommand{\ee}{\end{equation}}

\newcommand{\kms}{{\rm km\, s^{-1}}}

\newcommand{\msun}{{M_{\odot}}}
\newcommand{\msunyr}{{M_{\odot}}~yr^{-1}}

\begin{document}

\title{Evidence for a luminosity-class dichotomy in the supergiant donors of persistent high-mass
X-ray binaries and supergiant fast X-ray transients}

\titlerunning{Luminosity-class dichotomy in persistent HMXBs vs SFXTs}

   \author{L. Sidoli\inst{1}\fnmsep\thanks{Corresponding author: lara.sidoli@inaf.it}
       \and V. Sguera\inst{2}
        }

  \institute{INAF-IASF Milano, Istituto di Astrofisica Spaziale e Fisica Cosmica, via A. Corti 12, 20133 Milano, Italy
   \and INAF-OAS, Osservatorio di Astrofisica e Scienza dello Spazio, via Gobetti 101, 40129 Bologna, Italy}

   \date{Received May 18, 2026  /  Accepted August 11, 2026}

  \abstract
   {We have investigated the properties of the supergiant donors in two types of high-mass X-ray binaries
   (HMXBs, i.e. the so-called supergiant fast X-ray transients (SFXTs) and the classical X-ray sources),
   by collecting all available information from the literature.}
   {A prevalence of O-type versus B-type supergiant donors in SFXTs  was previously suggested in the literature, possibly
   indicative of faster winds in SFXTs than in classical, persistent sources.
   Our aim is to investigate whether there is evidence for a dichotomy in the stellar luminosity class, since B-type supergiants are also known to be hosted in SFXTs.}
   {Our collection of the relevant literature allowed us to build a robust sample of both types of HMXBs with a secure stellar classification.}
   {We have found that the SFXTs showing the most extreme, transient X-ray emission
   host only O-type or B-type stars with a Ib and Iab luminosity class, whereas classical sources have mostly Ia donors. }
   {This is robust evidence that might explain the two different behaviours of HMXBs at X-rays
   since the stellar luminosity class appears to be linked to the wind properties (the supergiant wind terminal velocity $v_\infty$,
   the mass-loss rate $\mdot_w$, and the wind strength $Q$ parameter),
   indicative of faster and less dense winds in SFXTs than in classical sources.}


   \maketitle

    \nolinenumbers

\section{Introduction}

High-mass X-ray binaries (HMXBs) are massive binaries typically hosting a neutron star (NS) that accretes a fraction of the wind
of the OB supergiant companion along the orbit \citep{Martinez2017, Kretschmar2019, Fornasini2023}.
They display two different X-ray behaviours: the classical wind-fed systems exhibiting persistent (although variable) X-ray emission (SgXBs)
and the transient sources (the so-called supergiant fast X-ray transients; SFXTs).

The SFXTs were discovered 20 years ago by the INTEGRAL satellite catching their bright  (L$_{X}$$\sim$10$^{36}$~erg~s$^{-1}$),
one-hour short X-ray flares \citep{Sguera2005, Sguera2006, Negueruela2006}.
These flares are typically part of very rare, few-days-long outbursts \citep{Romano2007},
separated by long intervals of much fainter X-ray emission (L$_{X}$$\lsim$10$^{34}$~erg~s$^{-1}$; \citealt{Sidoli2008}).
The percentage of time spent by SFXTs in outburst (L$_{X}$$\gsim$10$^{35}$~erg~s$^{-1}$) is less 5\%
and the dynamic range (i.e. the amplitude) spanned by their X-ray flux, from quiescence to the flares' peak,
is in the range 10$^2$-10$^6$ \citep{Sidoli2017}.

This behaviour indicates a fundamentally different coupling between the NS and the supergiant wind in SFXTs and in classical SgXBs \citep{Bozzo2008, Shakura2015}.
Whether this dichotomy arises from different properties  of the  NS (rotational velocity and surface magnetic field)
and/or of the supergiant wind (wind density and velocity at the NS's distance) is still much debated.
The NS surface magnetic field in SFXTs has been determined in a direct way (by means of cyclotron resonant scattering features, CRSFs) in only one source \citep{Bhalerao2015}, and is consistent with the ones measured in SgXBs.
A similar conclusion is suggested indirectly for other SFXTs by the empirical relation between the CRSF energy and the spectral cutoff energy \citep{Staubert2019}.
The pulsar spin periods in SFXTs, when known, largely overlap with the one shown by SgXBs.
The same is valid for the orbital geometry, altough SFXTs cover a broader range of orbital periods, reaching longer orbital periods,
up to 165 days \citep{Sidoli2007}.

The  alternative is to look at the specific properties of the wind outflowing from their supergiant companions, trying to highlight any systematic difference
between SFXTs and classical sources.
In fact, the supergiant wind properties play a crucial role in
accretion theory \citep{Stella1986, Bozzo2008, Shakura2015, Shakura2017, Martinez2017}: the accretion radius
($R_{acc}$, representing the gravitational reach of the compact object on surrounding matter)
is proportional to $v_{rel}^{-2}$, where  $v_{rel}$ is the relative velocity
of the NS with respect to the supergiant wind, often approximated with
the supergiant wind terminal velocity, $v_{\infty}$.
Moreover, the  accretion-driven X-ray luminosity is proportional to $\mdot_w$ $v_{rel}^{-4}$, according to the
simplest case of wind-fed  Bondi-Hoyle accretion, when applicable.

It has also been proposed that the magnetic field carried by the clumpy supergiant wind captured
by the NS might play a role in producing the SFXTs X-ray flares \citep{Shakura2014, Shakura2017, Hubrig2018}.
However, observational studies specifically addressing the properties of the winds of companion stars in  SFXTs
are still scarce in the literature \citep{Gimenez2016, Hainich2020, vandeneijnden2023, vandeneijnden2025}.

The investigation presented here follows a previous review paper on SFXTs \citep{Sidoli2017} and was also inspired by the review made by \citet{Negueruela2019}. In both studies, a possible tendency was highlighted for supergiant companions in SFXTs to be preferentially O spectral type rather than B-type (compared with  classical SgXBs), although the limited number of known systems prevented any statistically robust conclusion.
Since the winds from O-type supergiants  are typically faster than in B-type ones  \citep{Kudritzki2000},
this  would imply a less effective accretion in SFXTs, leaving more room for the onset of some gating mechanisms halting accretion in SFXTs, for most of the time.

More recently, observations performed at radio and millimetre wavelengths demonstrate that SFXTs are systematically underluminous at these wavelengths with respect to SgXBs \citep{vandeneijnden2025}, indicating systematically different supergiant wind properties  between the two types of HMXBs. Specifically, a lower ratio, $\mdot_w$/$v_{\infty}$, in SFXTs than in SgXBs.
This is particularly evident when comparing Vela X-1 (the prototypical system of classical, persistent SgXBs) and SAX~J1818.6-1703 (one of the most extreme SFXTs),
which share a similar distance and spectral type (B0.5), yet exhibit a markedly different flux density at 40--300 GHz
(which is proportional to ($\mdot_w$/$v_\infty$)$^{4/3}$),
implying less dense and/or faster winds in SFXTs than in SgXBs \citep{vandeneijnden2025}.

A potentially interesting feature that has not been highlighted before is at the origins of this paper: the B-type supergiant hosted in SAX\,J1818.6-1703 belongs
to luminosity class (LC) Iab \citep{Torrejon2010}, while the companion of Vela X-1 is classified as a Ia star \citep{MaizApellaniz2018}.
Early-type supergiants of LC Ia often display slower winds than Ib stars of a similar spectral type \citep{Kudritzki2000, Prinja1998}.
However, in these papers the average wind terminal velocity of supergiants with intermediate class Iab was not considered.
This is the starting point of the present investigation, which has the primary aim of
testing whether there is a prevalence of Ia sub-class in classical SgXBs with respect to SFXTs.
The secondary aim is to investigate if systematically different winds properties exist
in the different LCs (Ia, Iab and Ib).

The paper is organized as follows.
Sect.~\ref{sec:hmxbs} discusses the selection of our sample of spectrally identified HMXBs.
Sect.~\ref{sec:individual} outlines the results obtained from the literature about the spectral type and LC of single sources.
Sect.~\ref{sec:winds} discusses the supergiant wind properties of  our sample and, when this information
is not available in the literature, of other massive stars with the same spectral types and LCs (the `analogous stars’).
In Sect.~\ref{sec:discussion} we discuss the results and summarize the conclusions in Sect.~\ref{sec:concl}.

\section{Sample of HMXBs}
\label{sec:hmxbs}

We focus on Galactic HMXBs hosting a NS  or where a NS is strongly suspected to be the compact object,
i.e. we exclude the massive black hole binary Cyg\,X-1 from our investigation.
We note that all SFXTs are thought to host a NS, despite only a half of them having been confirmed as X-ray pulsars (e.g. \citet{Martinez2017} for a review).

The starting points for the selection of our sample of spectroscopically identified donor stars in both types of HXMBs
were the most recent catalogues of HXMBs \citep{Neumann2023, Kim2023, Fortin2023},
the papers focused on X-ray properties of SFXTs \citep{Sidoli2008, Sidoli2017, Romano2015, Sidoli2018, Romano2023},
and review papers on HMXBs \citep{Martinez2017, Sidoli2017, Negueruela2019, Kretschmar2019}.
We note that these papers and catalogues served only as an initial reference.
Each source was subsequently examined individually and the original ultraviolet (UV), optical and/or near-infrared (NIR) spectroscopic
study that classified the donor star was consulted (notes on individual objects are provided in Sect.~\ref{sec:individual}).

In a few cases we identified inconsistencies in the reported stellar type and/or LC,
sometimes accompanied by incorrect or misattributed references, or by classifications that were not strictly based on spectra of the optical counterparts.
For this reason, we preferred to carry out our own independent review of the literature.
In the end, we compiled our final sample exclusively from optically identified HMXBs (both classical SgXBs and SFXTs) that survived a comprehensive and rigorous, literature-wide spectroscopic screening,
retaining only systems with a secure UV, optical, or NIR spectroscopic classification providing a firm spectral type and (when available) LC (Ia, Iab, Ib).
The final sample is reported in Table\,B.1, with the relevant references.
The spectral types of the selected sample of supergiant HMXBs are the following: O6, O/WN9, O8, O8.5, O9, B0, B0.2, B0.5, B0.7, B1, B2, and B3, as listed in Table~B.1 (columns~5 and 6).

We rely on LCs derived from optical spectroscopy, whenever available,
as optical diagnostics are generally more sensitive to stellar LCs.
When only NIR spectra are available, different bands provide complementary constraints: H- and K-band spectra
primarily inform spectral type, while I-band features are more sensitive to gravity and better suited for luminosity classification \citep{Negueruela2008}.
K-band spectroscopy provides valuable complementary information; here, we use these results
only when optical classifications are not available.
Luminosity classes based solely on NIR spectra are therefore treated with appropriate caution.

When different spectral classifications are reported in the literature for the same source, generally
we retain the most sensitive results derived from optical spectroscopy.
In cases in which inconsistencies arise among published classifications,
we provide more details in the following dedicated subsections  about  individual sources (Sect.\,\ref{sec:individual}).

\begin{figure}[ht!]
    \centering
      {\includegraphics[width=6.2cm, angle=-90]{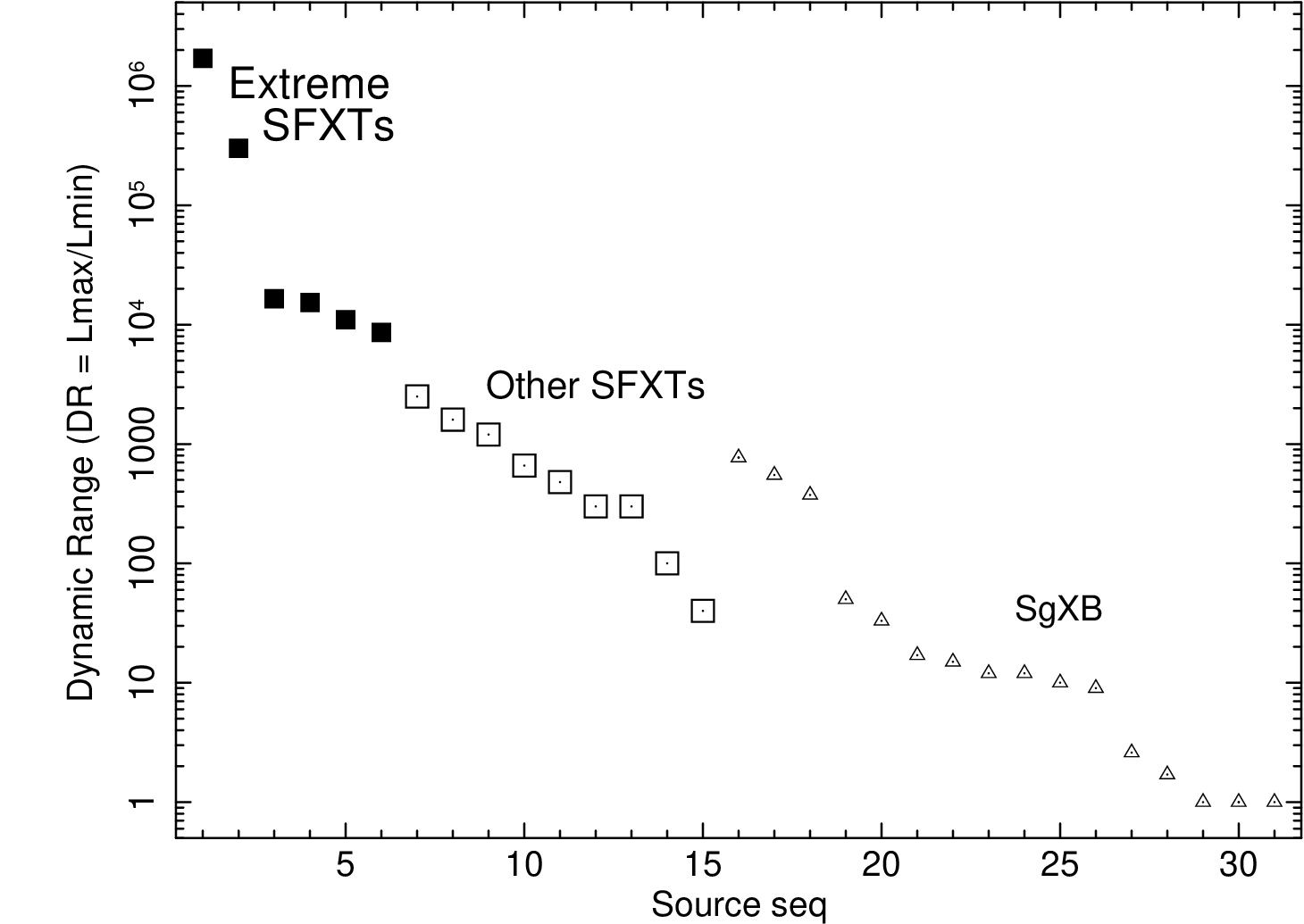}}
    \caption{Dynamic range (DR=max(L$_{X}$)/min(L$_{X}$)) reported in the literature (Table~B.1 for the data values and references) for our sample of 31 sources. Solid squares indicate `extreme SFXTs' (with DR$>$8000), open squares other SFXTs (with DR$<$8000), and open triangles SgXBs.
    }
      \label{fig:dr}
\end{figure}

In Table\,B.1, where our final sample of sources is reported,
we have distinguished SFXTs from SgXBs and, within the sample of SFXTs, we prioritize  SFXTs
displaying the most extreme transient X-ray emission, quantified by the greatest X-ray dynamic range
(`DR’; i.e. the ratio between the X-ray luminosity at the flare's peak and the lowest, quiescent X-ray luminosity).
In particular, the subset we designate as `extreme SFXTs’ includes only sources that are already unanimously classified as SFXTs in the literature, because of their high-DR values.
In order to clearly distinct this unambiguous sample of SFXTs, we adopted a literature‑guided threshold of DR$>$8000 (measured at soft X-rays, below 10 keV), solely to show that the bona fide SFXTs can be isolated through a simple quantitative parameter.
The values of DRs we have collected in Table\,B.1 (second column) from the literature are shown in Fig.~\ref{fig:dr} and
are mostly taken from  \citet{Sidoli2018} and \citet{Romano2023} (other references for DR values of a few sources
are listed in Table\,B.1, third column).
These most extreme SFXTs are unequivocally distinct from those of persistent, classical SgXBs.

The SFXTs showing a more limited (i.e. DR$\sim$100-1000) long-term X-ray flux  variability
are usually named `intermediate' SFXTs \citep{Rahoui2008igr18483}.
In a  few  intermediate cases there is no consensus in the literature
about their classification as SFXTs or as classical SgXBs.
For this reason, we have made this further, important selection.

We note that among the  SgXBs listed in Table\,B.1  there are also a few cases of an LC  with no correspondence
in the SFXT class, i.e. 4U~1700-37 (the only HMXB with an O6 supergiant),
GX~301-2 (with a hypergiant companion), and the peculiar OAO~1657-415 (with a O/WN9 fpe donor star).
There are also SgXBs with a quite large DR, overlapping with the one displayed by the intermediate SFXTs.
However, it is important to note that in  SgXBs a large DR can be merely due to the X-ray flux modulation along
an eccentric orbit (individual cases are discussed in Sect.\,\ref{sec:individual}).

\section{Results}
\label{sec:individual}

Our results for the HMXB sample (including both persistent and transient X‑ray sources) with secure spectroscopic classifications of their optical counterparts are summarized in Table\,B.1, with a special focus on the spectral type and the LC of the companion star.
In the subsections below, we provide brief notes only on those individual sources for which additional details are needed to justify the final classifications reported in Table\,B.1.

\subsection{Classical SgXBs}

The following brief subsections  concern only the classical systems  that require additional explanation.

\subsubsection{Vela X-1}

Vela X-1 is unanimously considered the prototypical source of the classical SgXBs showing
persistent X-ray emission  \citep{Martinez2017, Kretschmar2019, Kretschmar2021}.
The review by \citet{Kretschmar2021} provides a comprehensive overview of the
literature on this source and its various classifications, ultimately supporting the classification (B0.5Ia)
obtained in the GOSSS \citep{MaizApellaniz2018} and also listed in the SIMBAD database, at the time of writing.

The wind properties of its supergiant companion HD\,77581 have been directly measured by \citet{Gimenez2016}
($v_\infty$=700$^{+200}_{-100}$\,$\kms$; Table~B.1);
therefore, a comparison with wind velocity and mass loss rates from other analogous B0.5Ia stars is not strictly necessary.
We note, however, that \citet{Prinja1990} reports a wind terminal velocity of 1105\,$\kms$ for HD\,77581 and that
1700\,$\kms$ was reported by  \citet{Dupree1980}.
This latter result is discussed by \citet{Gimenez2016} (who observed 700\,$\kms$),
as both papers have some IUE spectra in common: \citet{Gimenez2016} conclude that this discrepancy is due to the fact that they
have taken into account  X-ray radiation in the line modelling
(i.e. X-rays intrinsic to the supergiant wind, not the X-ray emission from the Vela X-1 pulsar)
that was not accounted for by Dupree et al.
Here, we rely on the most recent measurement of the supergiant wind by \citet{Gimenez2016}.

\subsubsection{2S0114+65}

The spectral type and LC was obtained from optical spectroscopy \citep{Reig1996},
but we do not consider here the wind properties discussed in this paper, as they were not
directly measured through spectroscopy, but instead inferred from empirical scaling relations.

\subsubsection{4U~1538-52}

The optical spectrum of QV~Nor, the donor star in 4U~1538-52, has been examined in three papers.
\citet{Parkes1978} presented optical spectra in the range 3000-7000~\AA\ and classified the star as B0.2~Ia (as also reported in the SIMBAD database).
\citet{Crampton1978}, analysing independent optical spectra, noted that the observed features
were consistent with those of B0~Iab stars, although in both their conclusions and their summary table
they adopted the more general label B0~I without specifying a luminosity subclass.
\citet{Reynolds1992},  performing red-optical (6300-6700~\AA) spectroscopy,
concluded that QV~Nor is compatible with an early B-type supergiant, between B0~Ia and B1~Ib, based on the equivalent width of the
line He\,\textsc{i}\,$\lambda$6678\,\AA.
No subsequent optical, UV, or NIR spectroscopic study has refined these results, to the best of our knowledge.
Later works occasionally cite narrower subclasses (e.g. ‘Iab’ \citet{Neumann2023}\footnote{http://astro.uni-tuebingen.de/~xrbcat/ } at the time of writing),
without presenting new diagnostic evidence.
Given the spread among published classifications
the luminosity subclass of QV~Nor remains uncertain.
We note that it is reported simply as a B0I star in the review by \citet{Negueruela2019}.
Therefore, we adopted an early B supergiant (B0-B1), without assigning a specific luminosity subclass.

\subsubsection{IGR~J16207-5129}

The spectroscopy of its companion star was performed by \citet{Negueruela2007} and \citet{Nespoli2008}.
The first paper reports on the 6300-8900~\AA\ spectrum, concluding that the optical counterpart is
a B0 supergiant with an uncertainty of about one subtype.
\citet{Nespoli2008} analysed a K$_{s}$ spectrum, classifying this object as a B1~Ia star.

\subsubsection{IGR~J16320-4751 (aka AX J1631.9-4752)}

\citet{Coleiro2013} observed the NIR spectrum of this highly absorbed, obscured HMXB, classifying the counterpart as
a BN0.5Ia star, a classification we adopt here that is also consistent with the review by \citet{Negueruela2019}.
Earlier, \citet{Chaty2008} proposed an O-type supergiant based on their analysis of the NIR spectra,
particularly the red region.
The presence of a broad Br$\gamma$ line emission line led them to favour an O-type supergiant, in agreement with the
spectral energy distribution fitting \citep{Rahoui2008}.

\subsubsection{IGR~J16493-4348}

The only paper in which the spectroscopy (K$_{s}$ region) of its massive companion is reported
is \citet{Nespoli2010}, in which a spectral type B0.5-1~Ia-Ib was derived.
More recent classifications are not based on new spectroscopic data,
but instead on constraints from X-ray eclipse timing analyses
(e.g. B0.5~Ia suggested by \citealt{Pearlman2019}).
For this reason, we adopt only the spectroscopically derived classification B0.5-1~Ia-Ib.

\subsubsection{OAO~1657-415}

The NIR spectroscopy led \citet{Mason2009} to classify it as an Ofpe/WNL star,
implying a more evolved state than usual OB stars in HMXBs, in transition towards
hydrogen depleted Wolf–Rayet (WR) stars.
We note that, although this paper discusses an Ofpe/WNL classification, the K-band spectrum
is compared with the more specific subtype Ofpe/WN9 that has often been adopted in the literature.
It is worth noting that \citet{Mason2012} analysed spectroscopy in the H band,
finding that it is more similar to early B hypergiants
than the Ofpe/WNL classification inferred from the K-band observations alone.
However, these authors concluded that the correct identification is still a Ofpe/WN9 star.

\subsubsection{4U~1700-37}

This X-ray source hosts the most massive companion star in a HMXB, the bright star (V = 6.5 mag) HD~153919.
The literature about this star is vast.
For our purposes it is sufficient to note that its spectral type was determined in the 1970s
to be a O7f or O5.5f \citep{Jones1973, Walker1973}, later reported as O6.5~Iaf+ \citep{Prinja1990, Clark2002},
and subsequently refined to O6~Iafcp in more recent spectroscopic surveys (GOSSS; \citep{Sota2014}).
However, given the unusual character of this star and the fact that there are no SFXTs with
a similar identification, we shall not discuss this source further.

\subsubsection{IGR~J17252-3616 (aka EXO~1722-363)}

The donor star in EXO~1722-363 was classified as  a B0-B1 Ia star using a K-band spectrum \citep{Mason2009}.

\subsubsection{IGR~J18027-2016 (aka SAX~J1802.7-2017)}
\label{sec:sax1802}

This source is usually considered an SgXB \citep{Neumann2023, Negueruela2019}, but we note that
the range of X-ray flux variability in its long-term light curve exceeds two orders of magnitude \citep{Sidoli2018},
typical of intermediate SFXTs.

The most recently published NIR spectroscopy (I-band and K-band) of its companion star
resulted in the identification of a B1~Ib type \citep{Torrejon2010}.
However, the authors noted that the NIR spectroscopy was consistent with
both a B1~Iab and a B1~Ib classification, and that
distinguishing between the two LCs is challenging with the available spectral data.
Ultimately, they favoured the B1~Ib classification, as the relatively short orbital period of 4.6 days
appears incompatible with a B1~Iab star.

\subsubsection{XTE~J1855-026}

\citet{Negueruela2008xte1855} classified its donor star as a
B0 Iaep luminous supergiant from  high-quality spectra
taken with the 4.2-m WHT in La Palma.
Later, this classification was refined  to  BN0.2~Ia \citep{Negueruela2019}.

\subsubsection{4U~1907+097}

\citet{Cox2005} investigated the high-resolution  spectrum from 4680 to 10400~\AA, finally classifying it as an O8-9 supergiant, with
no conclusion about the specific LC (although we note that models for an O8Ia and an O9Ia stars have been adopted in this paper).
Later,  the analysis of K$_{s}$-band spectra yielded a O9.5~Iab classification  \citep{Nespoli2008}.
We note that the wind properties adopted in the discussion of this X-ray sources by \citet{Cox2005} are not directly measured from the spectrum
but assumed based on the stellar spectral type and on the scaling of the terminal wind velocity with the escape velocity
(i.e. $\mdot_w$=7$\times10^{-6}$~$\msunyr$ and v$_{\infty}$=1750~$\kms$).
Therefore we have not reported them in Table~B.1.

\subsubsection{4U1909+07 (aka X1908+075)}

\citet{Ge2024} present a sample of X-ray binaries, including their orbital parameters
and the spectral types of their optical counterparts.
For 4U1909+07, they report  an outdated spectral classification (O7.5-9.5I),
based on the catalogue of X-ray pulsars by \citet{Kim2023} and incorrectly attributed to \citet{Martinez2015}.
However, the classification as an O-type supergiant was originally published by \citet{Morel2005},
and later revised by \citet{Martinez2015}, who reclassified the optical counterpart as a B0-B3 star, likely a supergiant.
We also note that \citet{Negueruela2019} lists a B1-B3 I star and flags this as a secure classification,
attributing it to \citet{Martinez2015}.
If \citet{Martinez2015} is indeed the reference for this information,
we emphasize that \citet{Martinez2015} report a spectral type of B0-B3 star, not B1-B3.
Therefore, we consider it here to be a B0-B3 supergiant star.
The stellar wind properties were investigated by means of H- and K-band spectroscopy \citep{Martinez2015}
and confirmed with NOEMA observations at 100 GHz \citep{vandeneijnden2023}.

\subsubsection{IGR~J19140+0951}

Spectroscopic studies in the H and K bands identified
the counterpart as a B0.5 supergiant \citep{Hannikainen2007}, while
a B1~Iab identification based on a K$_{s}$-band spectrum was reported by \citet{Nespoli2008}.
I-, J-, H-, and K-band spectra confirmed a B0.5 supergiant but with a higher luminosity (B0.5~Ia; \citet{Torrejon2010}).
The large range of variability in its long-term X-ray flux (almost three orders of magnitude, more typical of SFXTs)
was pointed out \citep{Sidoli2016},  possibly suggestive of a smooth transition in-between classical SgXBs and SFXTs.

\subsection{Extreme SFXTs}

Only  extreme SFXTs demanding an extra discussion on their donor stars
are briefly noted below.

\subsubsection{IGR~J17544-2619}

The optical and NIR  spectroscopy covering the range of wavelengths 4000-8000~\AA\ enabled the identification of an O9Ib star \citep{Pellizza2006}.
Later, \citet{Gimenez2016} confirmed the spectral type by means of optical and NIR spectroscpy, refined the source distance
to 3.0$\pm{0.2}$~kpc, and estimated for the first time the properties of its supergiant wind
($v_{\infty}$=1500$\pm{200}$~km~s$^{-1}$ and $\mdot$=1.6$^{+0.9} _{-0.6}$$\times10^{-6}$~$\msun$/yr).
These authors concluded that the optical and NIR
spectra of this SFXT showed nothing peculiar with respect to other  O9I-type stars.

\subsubsection{XTE~J1739-302 (aka IGR~J17391-3021)}

Optical and infrared spectroscopy  identified its massive donor as an O8~Iab(f) star \citep{Negueruela2006xte1739}.
As concerns its long-term X-ray behaviour, the lowest luminosity state
was captured with \xmm\ and reported by \citet{Sidoli2023xte1739}, updating its dynamic range to five orders of magnitude,
reaching the most extreme SFXTs, as reported in Table~B.1.

\subsubsection{IGR~J16479-4514}

The companion star was always spectrally classified as an O-type supergiant
(O9.5Iab, \citet{Nespoli2008}; O8.5Ib \citet{Negueruela2019}).

\subsubsection{AX~J1841.0-0536 (aka IGR~J18410-0535)}

Spectroscopic analysis of the donor star in the K$_{s}$ band yielded a B1~Ib classification  \citep{Nespoli2008},
whereas optical and infrared observations led to a B0.2~Ibp classification \citep{Negueruela2008, Negueruela2019}.
Given the broader wavelength coverage of the latter studies, crucially including optical spectra,
we adopt the classification of the supergiant star derived from those results.

\subsubsection{SAX~J1818.6-1703}

The analysis of the K-band spectrum of the IR counterpart resulted in the determination of a B0-B1.5 supergiant,
with a precise luminosity sub-class being difficult to assign \citep{Torrejon2010}.
Nevertheless, these authors conclude that it is `likely to be at least Iab’, placing the source at
a distance of 2.1$\pm{0.1}$~kpc, in the outer regions of the Sagittarius arm.

\subsection{Others SFXTs}

We give here some additional comments on SFXTs that show a smaller amplitude of X-ray flux variability
and are named `other SFXTs' in Table~B.1.

\subsubsection{IGR~J00370+6122}

This source is listed among SFXTs by \citet{Negueruela2019} and considered an intermediate SFXT by \citet{Hainich2020}.
Since it is a source less investigated than other HMXBs, we extracted a \sw/XRT (0.3-10 keV) long-term light curve
using the pipeline made publicly available by the UK Swift Science Data Centre \citep{Evans2007, Evans2009},
where it shows a dynamic range of $\sim$300 (Fig.~\ref{fig:xrtlc}).
This amplitude of variability is indeed consistent with an intermediate SFXT.

\subsubsection{IGR\,J11215-5952}

The source IGR\,J11215--5952 is peculiar, being the only SFXT
showing strictly periodic X-ray outbursts, every 165 days \citep{Sidoli2006, Sidoli2007, Romano2007, Sidoli2017igr11215}.
This periodicity is interpreted as the orbital period of the system  \citep{Sidoli2007},
very similar to type-I outbursts of Be/XRBs \citep{Reig2011}.
As remarked by \citet{Negueruela2019}, the large dynamic range might be produced only by the very eccentric orbit.

\subsubsection{IGR~J16195-4945 (aka AX~J161929-4945)}

\begin{figure}[ht!]
    \centering
    {\includegraphics[width=6cm, angle=-90]{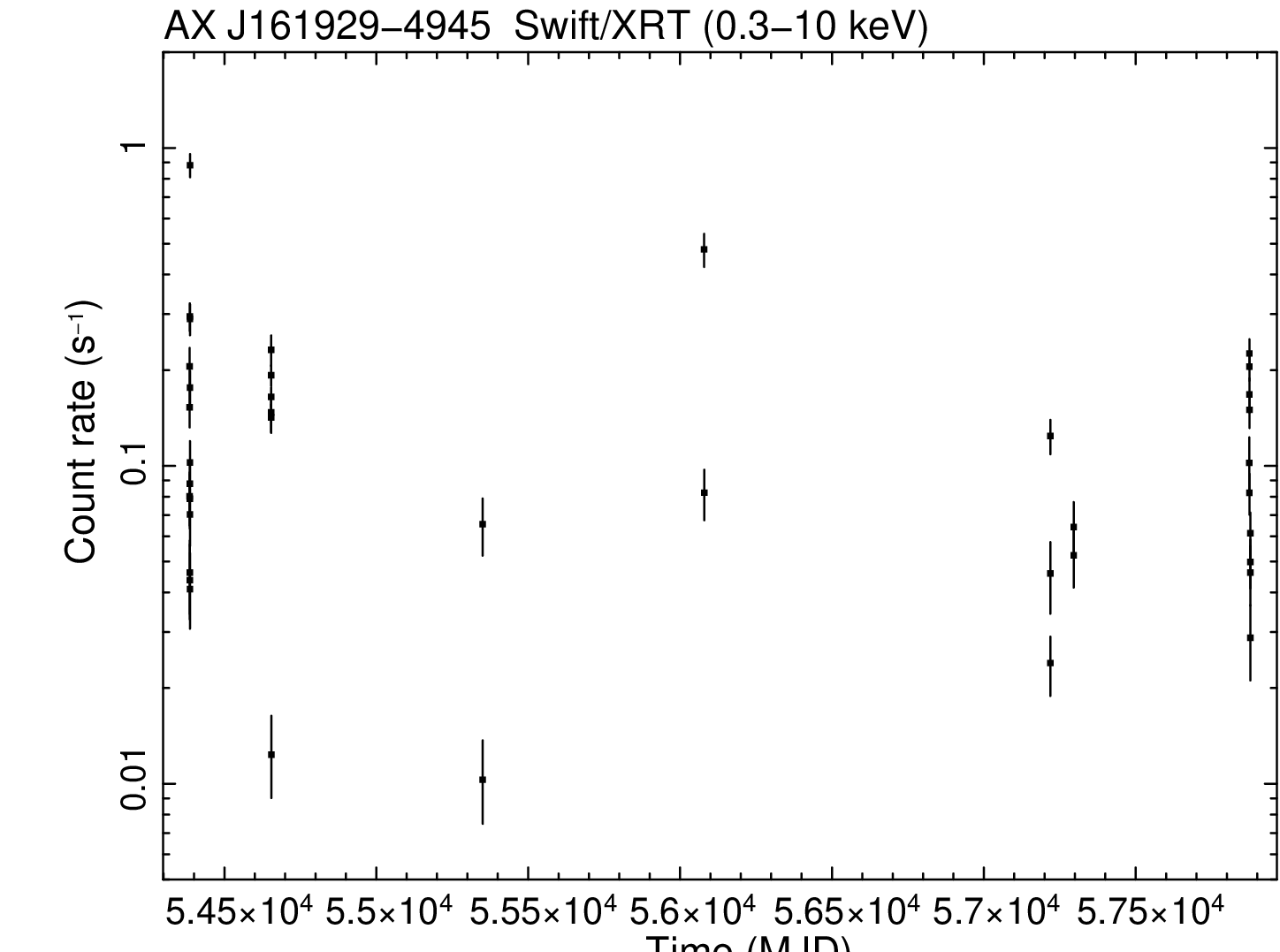}} \\ \vspace{6mm}
       {\includegraphics[width=6cm, angle=-90]{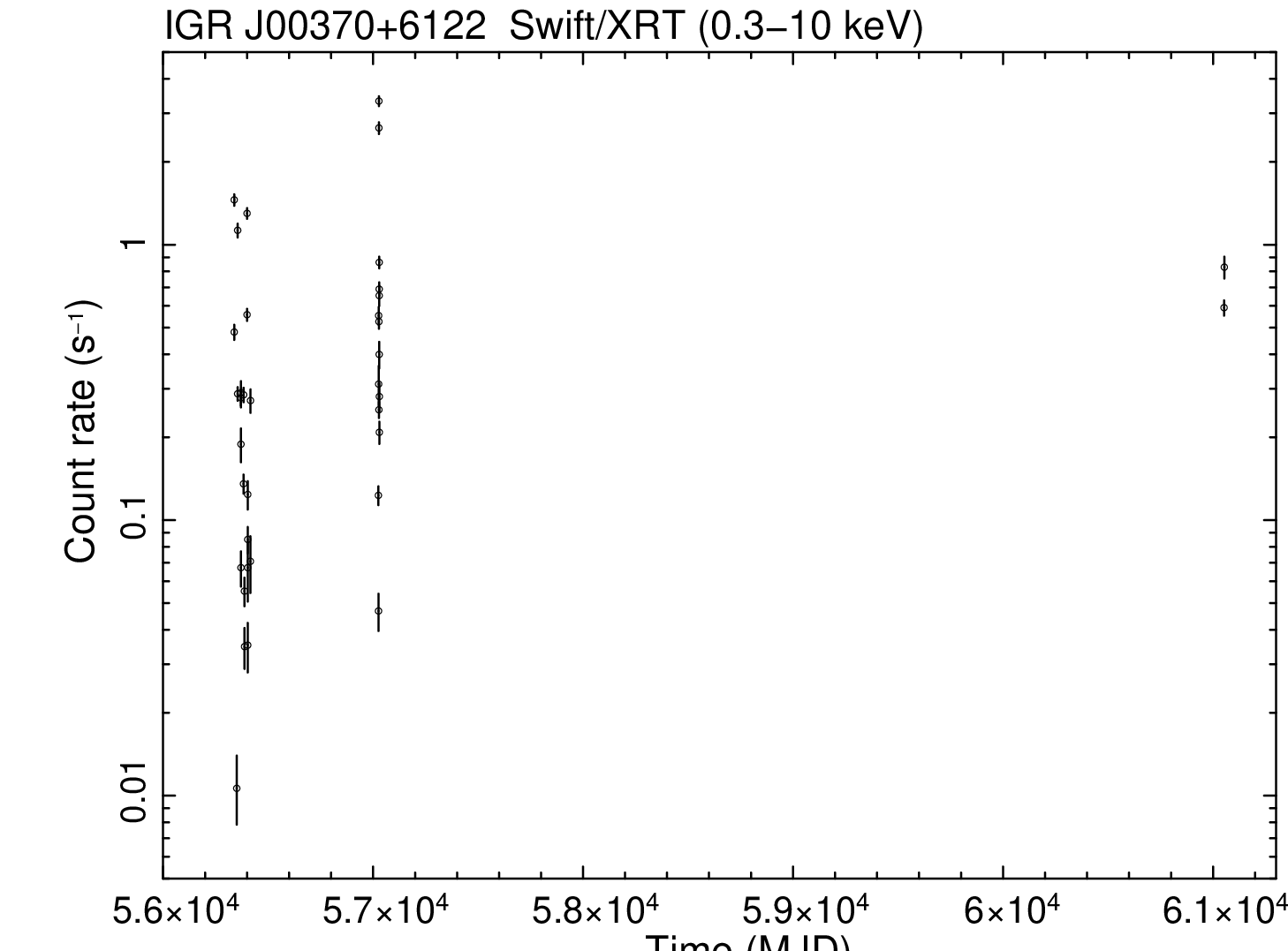}} \vspace{4mm}
    \caption{Long-term X-ray light curves (\sw/XRT; 0.3-10 keV) of AX~J161929-4945 (upper panel) and IGR~J00370+6122 (lower panel) that
    we extracted making use of the UK Swift Science data Centre \citep{Evans2007}.}
      \label{fig:xrtlc}
\end{figure}

IGR~J16195$-$4945 is occasionally referred to in the literature as either a candidate SFXT \citep{Satybaldiev2023} or a persistent, classical SgXB \citep{Kretschmar2019, Neumann2023}.
The most recent ART-XC light curve shows only a modest change between the quiescent and active phases (merely a factor of $\sim5$), from $9\times10^{-12}$ to $4.1\times10^{-11}\,\mathrm{erg\,s^{-1}\,cm^{-2}}$, far below the large amplitudes (typically $\gtrsim 10^{2}$) that characterise bona fide SFXTs \citep{Sidoli2017}.
However, the \sw/XRT (0.3-10 keV) long-term light curve we extracted using the pipeline made publicly available
by the UK Swift Science Data Centre\footnote{https://www.swift.ac.uk/xrt\_products/ \citep{Evans2007, Evans2009}}
shows a dynamic range of about 2 orders of magnitude (Fig.~\ref{fig:xrtlc}).
This is more consistent with an intermediate SFXT than with a SgXB,
according to the commonly adopted classification scheme \citep{Sidoli2017}.

The NIR counterpart 2MASS J16193220-4944305 (also reported in the GLIMPSE catalogue as G333.5571+00.3390; \citet{Tomsick2006})
was spectrally classified as an ON9.7Iab star \citep{Coleiro2013} in their
analysis of  H and K$_S$ spectra.
No further papers have been published on this optical counterpart.

\subsubsection{IGR~J16328-4726}

\citet{Coleiro2013} classified it as an O8Iaf (or more probably an O8Iafpe) star in their
investigation of the K$_{s}$~spectrum.

\subsubsection{IGR~J16418-4532}

The companion star of IGR~J16418-4532 is classified as a BN0.5Ia star in the SIMBAD database,
based on the K$_{s}$-band spectroscopy performed by \citet{Coleiro2013}.
An alternative classification (O9.5I star) is reported by \citet{Goldoni2012}, who analysed
the X-Shooter spectrum in the same NIR band.
The source was too faint to obtain a reliable spectrum at shorter wavelengths.
These different spectra suggest a variable stellar wind and/or circumstellar environment in
a late O-type companion in IGR~J16418-4532.

\subsubsection{IGR~J16465-4507}

Its optical counterpart has been classified as either a B-type supergiant
(B0.5-B1Ib, \citealt{Goldoni2012,Chaty2016}, refined to B0.5Ibn by \citealt{Negueruela2019})
or a O9.5Ia star \cite{Nespoli2008}.
We follow the most recent B-type classification, which is grounded in optical/UV diagnostics of the O9.5-B0.5 boundary,
unlike the earlier NIR-based O9.5 Ia identification, which was based on a spectral analysis restricted to the K band.

\subsubsection{IGR~J17354-3255}

The K-band spectroscopy led to the identification of an O-type supergiant, somewhere in between
O8.5Iab(f) and  O9Iab spectral types \citep{Coleiro2013}.

\subsubsection{AX~J1845.0-0433 (aka IGR~J18450-0435)}

The most recent optical to infrared spectroscopy identified it as an O9~Ia star  \citep{Negueruela2008, Negueruela2019}.
A conclusion consistent with these results (O9.5~I, with a range of  $\pm{0.5}$ on the spectral classification)
was also obtained by \citet{Coe1996} by means of optical spectroscopy.

\subsubsection{IGR~J18483-0311}

The companion star of this source was classified at first as ``probably” being a “B0.5~Ia star” based on  H- and K-band spectroscopy
\citep{Rahoui2008igr18483}, and as a B0.5-1~Iab star based on I-, H-, and K-band spectra \citep{Torrejon2010}.
In particular, the detection of Paschen lines only up to P18 suggests that the star is not a Ia star  \citep{Torrejon2010}.

\section{Results for stellar wind parameters and `analogous' stars}
\label{sec:winds}

Once we had selected our sample of optically identified HMXBs, we surveyed the literature for   spectroscopic studies
aimed at directly determining their wind properties, i.e. the wind terminal velocity ($v_\infty$)
and the supergiant mass loss rate ($\mdot_w$).
When available, we list them in the last columns in Table~B.1, as well as the relevant literature.
It is evident that the supergiant wind properties are known for only a few HMXBs.

We note that the recent radio/millimetre survey of a sample of HMXBs \citep{vandeneijnden2023, vandeneijnden2025}
has crucially highlighted a systematically lower emission at these wavelengths in SFXTs than in SgXBs
(and, consequently, a lower ratio, $\mdot_w/v_\infty$).
However, since it is not possible to disentangle $\mdot_w$ from v$_\infty$, we have not reported these wind results in Table~B.1.

For the great majority of HMXBs missing direct measurements, we decided to compare them with a sample of `analogous' stars;
namely, supergiants of the same spectral type and LC of our HMXBs' donors, for which the wind properties are known.
For Galactic OB supergiants, the terminal velocity and the mass loss rate are mostly listed in a few papers that date back to the 1980s and 1990s
\citep{PH86, HP89, Prinja1990}, which are still key references for most subsequent studies of massive stars,
especially for the wind terminal velocity derived from UV measurements \citep{deburgosletter2024}.

The terminal velocities reported in these earlier catalogues can be considered reliable quantities overall, whereas the mass loss rates are more uncertain: wind clumping
(i.e. the presence of density inhomogeneities throughout the outflow)
significantly affects the mass-loss diagnostics, leading
to a downward revision of previously estimated mass-loss rates \citep{Puls2006, Puls2008}.
Therefore we do not discuss mass loss rates reported in the literature in detail.

Clumpiness of supergiant winds  has also been suggested to play a role in driving the X-ray flaring  behaviour in SFXTs
(e.g. \citealt{Negueruela2008}, \citealt{Martinez2017} for a review).
On the other hand, constraining the clumpiness in stellar winds of massive stars is inherently difficult \citep{Hawcroft2024}
and the clumping factors derived from sensitive, high-resolution spectra of supergiant winds in SgXBs
show no significant differences compared to those obtained for SFXTs \citep{Hainich2020}.
For this reason, we do not discuss wind clumpiness further.

More recently, new spectroscopic surveys of massive stars have been performed, notably
the Galactic O-Star Spectroscopic Survey\footnote{https://gosc.cab.inta-csic.es/}
(GOSSS; \citealt{MaizApellaniz2011, MaizApellaniz2013, MaizApellaniz2018})
and the
IACOB project\footnote{https://research.iac.es/proyecto/iacob/pages/en/introduction.php}
(Instituto de Astrof\'isica de Canarias OB stars; \citealt{deburgos2023, Negueruela2024, deburgosletter2024, deburgos2024a, holgado2025}).
We used these surveys to select the sample of analogous stars, in order to adopt the most up-to-date spectral type and LC classifications (Table~C.1).

We next searched the literature for the wind properties of the selected stars, as the only quantity related with the supergiant winds
is the so-called `wind-strength’ $Q$ parameter listed by \citet{deburgosletter2024}.
The parameter \(Q\) for massive stars \citep{Puls1996} is defined as $Q$ = $\mdot_w / (R_{\star}v_{\infty})^{1.5}$  \citep{deburgosletter2024},
where $\mdot_w$ is the wind mass-loss rate (in solar masses per year),
$v_\infty$ is in units of kilometres per second, and
$R_{\star}$ is the stellar radius (in units of solar radii).
$Q$ is derived from the spectra of OB stars and provides a measure of the overall wind strength
in a single observationally constrained quantity that enables a direct comparison of
wind strengths across stars of different radii and wind speeds.

Moreover, we selected stellar and wind parameters of other massive stars, searching the literature,
to obtain as large as possible a sample of early-type supergiants
with similar spectral types and LC to the donor stars in SgXBs and SFXTs.
When the massive star is not listed in \citet{deburgos2024a}, we adopt the classification reported by \citet{Prinja1990}.
Sometimes the stellar classification differs in different papers; therefore, for each star we justify our  assumptions in Sect.\ref{sec:notes}.

The final sample of analogous stars is reported in Table~C.1.
For this sample, we calculated the average values of the wind properties, which we report in Table~C.2, for clarity.
We note that, beside this sample of analogous stars for which we found in the literature a measurement of the terminal velocity and/or of the mass loss rate,
we retain in Table~D.1 the analogous stars selected only from  \citet{deburgos2024a}, where the `wind-strength’ $Q$ parameter
is listed. We note that these two tables share some stars, but not all.
We also adopted this sample with a measured wind strength, because it includes 135 stars (of a total of 527 stars in the IACOB survey),
with spectral types from 09 to B3, allowing us to enlarge the sample  for a meaningful comparison with the companion stars in our sample of HMXBs.

We remark that in our investigation all wind parameters are drawn  from papers reporting primary spectroscopic measurements.
The sole exception is \citet{Markova2004}, who derived some values of the wind terminal velocity
by interpolating estimates from earlier studies; because those earlier sources are not otherwise adopted here,
the Markova et al. values are included without introducing redundancy.
We do not report here wind parameters reported by  \citet{Markova2004} for the O-type stars
HD\,338926, BD+56\,739, and HD\,18409, as they used calibrations provided by \citet{Kudritzki2000}.

\section{Discussion}
\label{sec:discussion}

\begin{figure}[ht!]
    \centering
      {\includegraphics[width=8.5cm, angle=0]{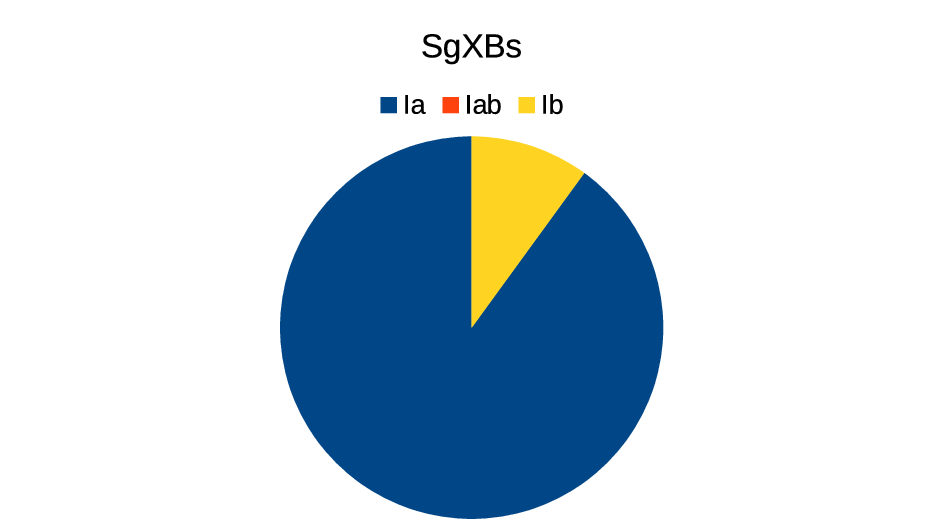}} \\
      {\includegraphics[width=8.5cm, angle=0]{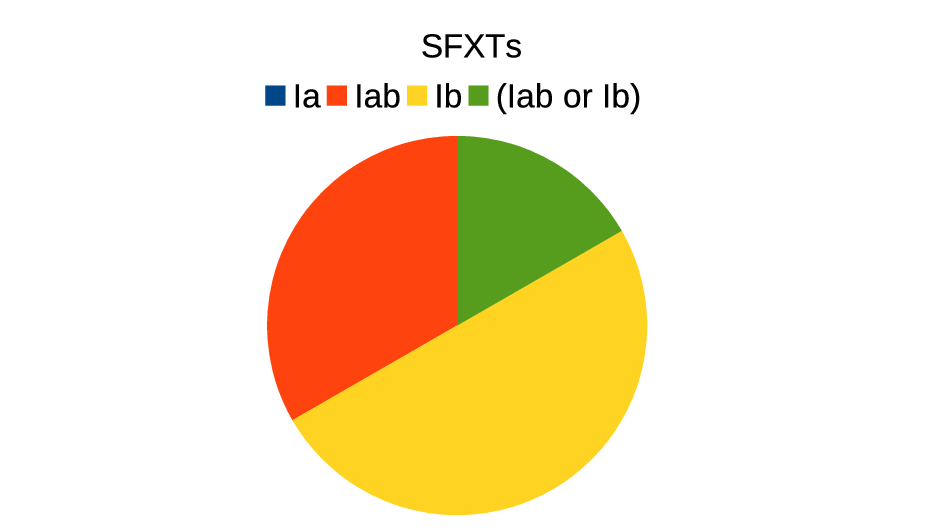}}
    \caption{Luminosity class (Ia, Iab, Ib) composition of early-type donor stars in the
    sample of classical SgXBs (upper panel)
    and of the most extreme SFXTs (lower panel), as reported in Table~B.1.}
      \label{fig:pies}
\end{figure}

The first robust result of our investigation is evident from Table~B.1, where
we have collected the classification of the optical counterparts of both kinds of HMXBs, taken from the literature:
the most extreme SFXTs (IGR~J17544-2619, XTE~J1739-302, IGR~J08408-4503, IGRJ16479-4514, SAXJ1818.6-1703, and AX~J1841.0-0536),
showing the largest variability in their flaring X-ray activity,
host supergiants  with Ib and Iab LCs (either O- or B-type stars).
It is remarkable that no Ia stars are present in this group of well-established SFXTs, while the Ia class dominates among SgXBs.
This dichotomy is shown in Fig.~\ref{fig:pies}.

O-type supergiants are  only present in 3 of 16 SgXBs. These three sources have peculiar properties:
4U~1700-37 and OAO~1657-415 have peculiar donor stars, while the third SgXB (4U~1907+097, hosting an O8-O9I star)
was defined as a missing link between the SFXTs and classical accreting pulsars \citep{Doroshenko2012}.

Among all early B-type supergiants (13 of 16 sources) dominating the persistent group of sources,
Ia class is well established in 9 SgXBs (of a total of 16 objects).
For the remaining 7 SgXBs, 4 of them have an unknown LC, 2 sources have a controversial one (IGR~J16207-5129 and IGRJ16493-4348),
and the SgXB OAO~1657-415 is peculiar, as the companion star has a completely different nature (O/WN9fpe).
Remarkably, there is only one B-type~Ib star among SgXBs: the B1~Ib star in SAX~J1802.7-2017; however, it is unclear whether it is
a persistent SgXB or an intermediate SFXT, as the amplitude of the variability of its X-ray intensity overlaps with the range of intermediate SFXTs \citep{Sidoli2008}.

\begin{figure*}[ht!]
    \centering
      {\includegraphics[width=15cm, angle=0]{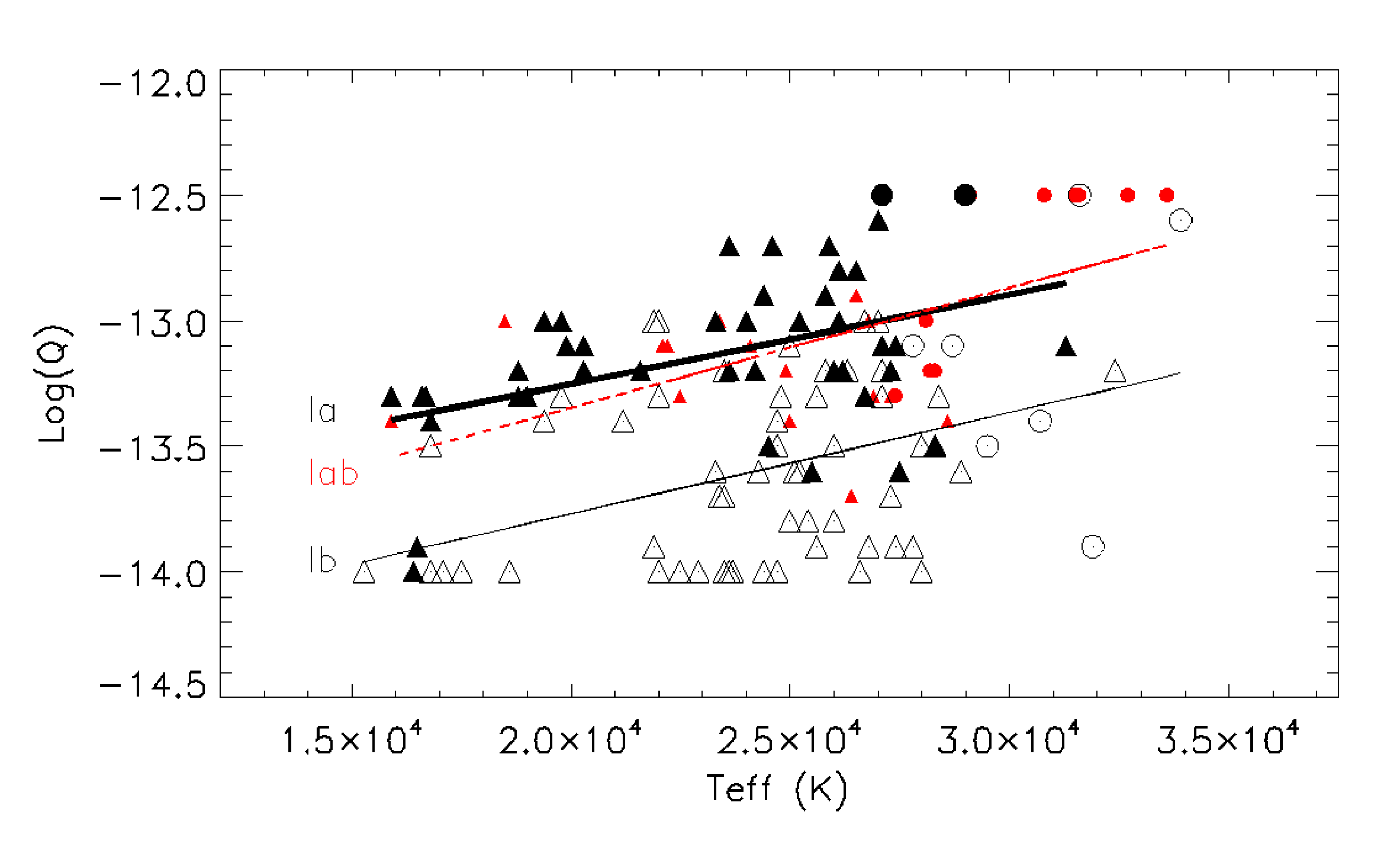}}
    \caption{Wind-strength parameter, $Q$ (expressed as $\log Q$), plotted as a function of the effective temperature,
    using stellar data selected from \citet{deburgos2024a} and reported in Table~D.1.
    O-type supergiants are plotted as circles, while  B-type supergiants appear as triangles.
    LC Ia stars are shown with filled black symbols,  Iab stars with red symbols,  and Ib stars with open black symbols.
    Uncertainties are mostly comparable with symbol sizes.
    To guide the eye, linear fits are overplotted: the thick black line corresponds to the best fit for OB~Ia stars,
    the red line to OB~Iab stars, and the thin black line  to the OB~Ib stars. Despite the significant scatter,
    it is evident that stars with Ib LC display weaker winds than Ia stars.}
      \label{fig:teffqlog}
\end{figure*}

\begin{figure}[ht!]
    \centering
      {\includegraphics[width=9.0cm, angle=0]{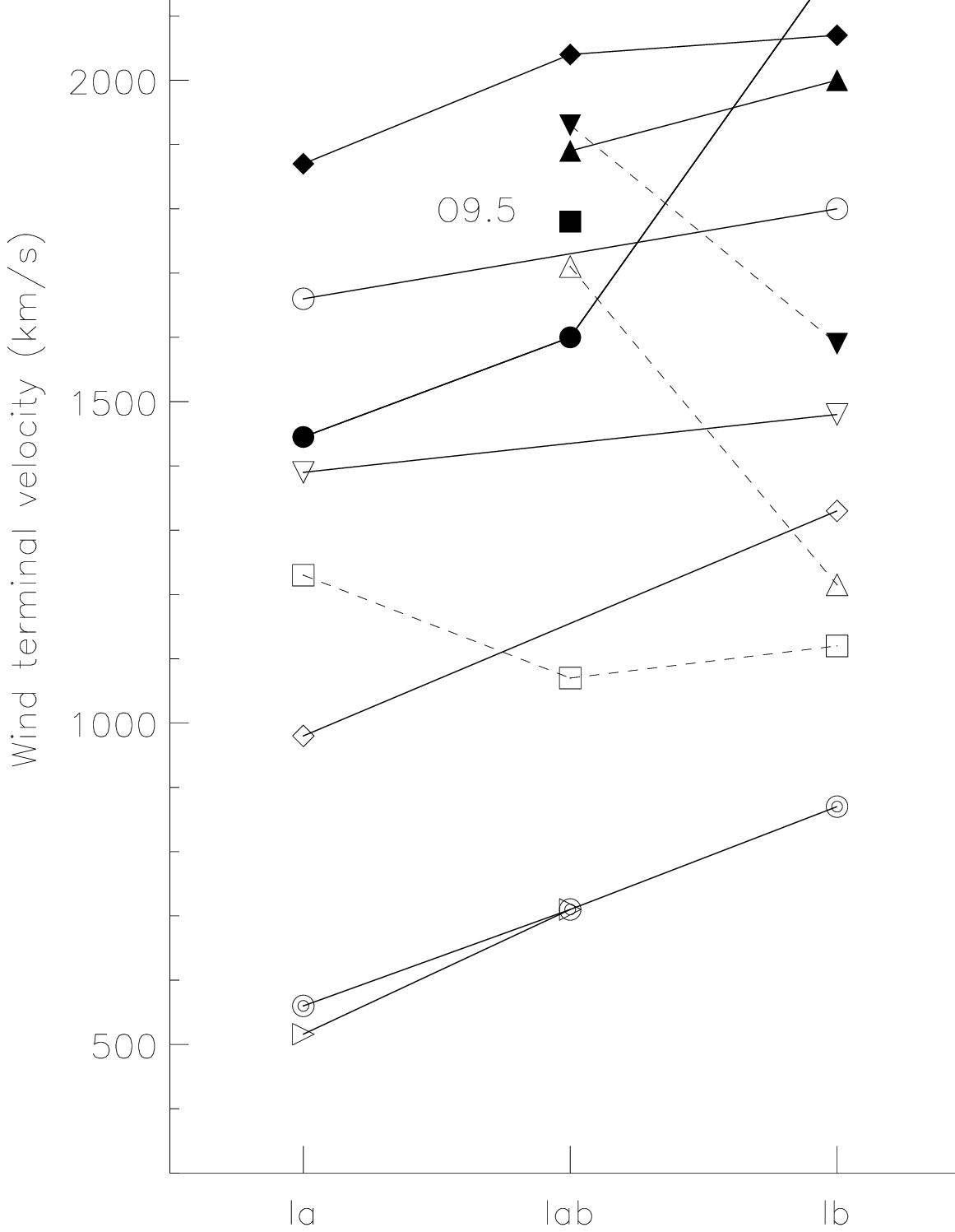}}
    \caption{Average wind terminal velocities versus LCs
    for the corresponding stellar analogs, for each spectral type marked
    on the right (values are taken from Table~C.2, third column).
    Solid symbols indicate O-type supergiants, while empty ones mark B-type stars.
    Solid lines connect increasing wind velocities with LC (from Ia to Ib),
    while dashed lines
    indicate decreasing or non-monotonic trends of the average wind terminal velocities with LC (O9.7, B0.2, and B1 stars, as discussed in Sect.~\ref{sec:discussion}).
    }
      \label{fig:vinflc}
\end{figure}

To quantify the statistical significance of the apparent
segregation in LC between classical SgXBs
and extreme SFXTs, we applied a Fisher exact test
\citep{Fisher1922,WallJenkins2012}. Given the limited sample
size and the presence
of contingency-table cells with very small numbers (including
zero counts), the Fisher exact test is more appropriate than
a standard $\chi^2$ test because it provides an exact probability
without relying on large-sample approximations.
We grouped the donor stars into two categories: Ia and non-Ia
(including Iab and Ib LCs).
For the first test, we considered the contingency table extracted
from Table~B.1, including the full sample of extreme SFXTs and
classical SgXBs with a secure LC s determination.
We excluded intermediate SFXTs (9 sources) that do not satisfy
the criterion for extreme dynamic range, DR$>$8000, as well as
classical SgXBs with ambiguous (2 sources), unknown (3 sources), or
peculiar (1 source) luminosity-class classifications.
Notably, only
one SgXB (SAX~J1802.7$-$2017) hosts a non-Ia
donor star, classified as a Ib:
\[
\begin{array}{c|cc}
& \mathrm{Ia} & \mathrm{non\mbox{-}Ia} \\
\hline
\mathrm{classical\ SgXBs} & 9 & 1 \\
\mathrm{extreme\ SFXTs} & 0 & 6 \\
\end{array}
.\]
The Fisher exact test yields a p value of $p = 8.7 \times
10^{-4}$, demonstrating that the observed
segregation is unlikely to arise by chance.
We then performed a second, more conservative test excluding
the only non-Ia classical SgXB (SAX~J1802.7$-$2017).
This source can be excluded from the SgXB sample because i) its LC, as reported in the literature \citep{Torrejon2010},
lies between Iab and Ib (see Sect.~\ref{sec:sax1802}),
and the available spectra do not allow these authors to make a firm distinction between these subclasses.
The Ib classification ultimately adopted by  \citet{Torrejon2010}
was  based solely on the short orbital period, which favours a smaller Ib donor over a larger-radius Iab star.
Moreover,   ii) its classification as a genuinely SgXB
is also somewhat ambiguous, since its
X-ray variability partially overlaps with the behaviour typically
observed in intermediate SFXTs (Sect.~\ref{sec:sax1802}).
In this case, the contingency table becomes

\[
\begin{array}{c|cc}
& \mathrm{Ia} & \mathrm{non\mbox{-}Ia} \\
\hline
\mathrm{classical\ SgXBs} & 9 & 0 \\
\mathrm{extreme\ SFXTs} & 0 & 6 \\
\end{array}
.\]
For this stricter selection, the Fisher exact test gives a p value
of
$p = 2.0 \times 10^{-4}$ further strengthening the significance of
the observed dichotomy.
These results quantitatively support the conclusion that extreme
SFXTs preferentially host Ib and Iab supergiant donors, whereas
classical SgXBs are dominated by Ia supergiants.
This observational segregation strongly suggests that the
different X-ray behaviours of the two classes are linked to
systematic differences in the stellar wind
properties associated with the LC of the donor
stars.

In fact, we next investigated the properties of the supergiant winds in our sample of HMXBs.
Firstly, only a few HMXBs were directly investigated through UV spectroscopy to determine their wind parameters (Table~B.1).
This scarcity largely reflects the substantial line-of-sight absorption, often local to the system, which severely limits the
feasibility of such observations with current instrumentation.
Nevertheless, the available spectroscopic studies show that the two extreme SFXTs observed (IGR~J17544-2619 and IGR~J08408-4503)
exhibit significantly faster winds than the three SgXBs observed (again, with the exception of the peculiar system SgXB 4U1700-37).
This difference is readily understood, as the SFXTs host O-type supergiants, whereas these three SgXBs (Vela X-1, 4U~1909+07 and GX~301-2) have B-type donors.
This fact, coupled with the results of the recent radio and millimetre surveys \citep{vandeneijnden2023, vandeneijnden2025}
confirm that faster (and/or less dense) winds characterise SFXTs, compared with SgXBs.

Our second result is that, when B-type supergiants are present in both types of HXMBs, the fact that only Iab and Ib stars are hosted in prototypical SFXTs
is highly suggestive: this observational pattern again points towards
less dense winds in SFXTs compared to classical SgXBs.
In fact, the wind strength ($Q$ parameter) of the Galactic OB stars we retrieved from the stellar database of the IACOB project (Fig.~\ref{fig:teffqlog}; \citet{deburgos2024a})
demonstrates that Ia stars of the spectral types of HMXBs donors
show systematically larger $Q$ values than Ib LC (although with some overlap), for each spectral type (or effective temperature; Fig.~\ref{fig:teffqlog}).
Iab stars mostly occupy a region in between Ia and Ib in this plane.
This dichotomy in the wind strength ($Q$) parameter (especially for B-type stars) is indicative
of denser winds (in the recombination-line forming region)
in Ia than in Iab and in Ib stars. This of course reflects, by the definition of $Q$, the combined
effect of a higher mass loss rate, a lower terminal velocity, a larger stellar radius, or any combination of these in Ia stars.
However, it is important to note that de Burgos et al. consider only homogeneous winds (i.e. without clumping);
therefore, the $Q$ values provide an upper limit for the wind strength.

A third piece of evidence becomes apparent when considering the data that we compiled from the literature about the analogous stars
and their average wind terminal velocities reported in Table\,C.2.
A direct trend of increasing average wind terminal velocities appears from Ia to Iab to Ib LCs for the following
spectral types: O8, O8.5, O9, B0, B0.5, B0.7, B2, and B3 (Fig.~\ref{fig:vinflc}).
However, the speed ranges of LCs always overlap for a given spectral type.
For the other spectral types, a trend is not observed in Table\,C.2 and in Fig.~\ref{fig:vinflc}.
An exception is the O9.5 spectral type,
for which we could only find terminal velocities in the literature for O9.5Iab stars.

All the information about OB supergiants collected here indicates larger terminal wind velocities in the extreme SFXTs than in SgXBs.
Since the accretion (Bondi) radius strongly depends on the velocity of the supergiant wind in the vicinity of the NS,
and since the NS magnetospheric radius depends on the density of the gravitationally captured wind material,
higher wind speeds and lower densities
make the NS more prone to transit to regimes where the accretion onto its surface is inhibited (e.g. \citet{Bozzo2008}).
Obviously the supergiant wind properties are not the only ingredient to explain SFXT behaviour, but we remark that what we have found,
i.e. that persistent versus SFXTs are significantly different in terms of the LCs of their supergiant companions, is a robust and crucial observational fact.

\section{Conclusions}
\label{sec:concl}

We have revisited the properties of the supergiant companions in both types of supergiant massive X-ray binaries, focusing especially on the role of the LC (Ia, Iab, and Ib), beyond spectral type alone (which has already been investigated in previous literature).
Our review of the spectral identification of the optical counterparts in SgXBs and SFXTs reveals that
there is evidence for a dichotomy in their LC, i.e. a significant prevalence of LC Ia in classical sources,
while among the confirmed SFXTs displaying the most
extreme amplitude of their variability in X-ray flux, the LC is only Ib and Iab.

From the comparison with analogous Galactic OB stars of the same spectral type and LC,
we have also found an indication of a dichotomy in the wind strength ($Q$ parameter), with larger values in Ia stars than Ib ones (although with some overlapping values),
indicative of denser winds in the line-forming region in Ia than in Iab and in Ib stars, independent of the specific values of $\mdot_w$, $v_\infty$ and stellar radius.
Our conclusion is that there is nothing really `peculiar’ in supergiant donors belonging to SFXTs,
but their  lower ratio, $\dot{M}$/$v_\infty$, with respect to persistent sources  might simply reflect their different
optical, stellar LC.
There is a clear need for future UV spectroscopic observations and new radio surveys to directly measure the wind parameters of
the majority of SFXTs (and SgXBs) for which such observations are still lacking.

\section*{Data Availability}

Tables B.1, C.1, C.2, and D.1 are available at the CDS via
https://cdsarc.cds.unistra.fr/viz-bin/cat/J/A+A/vol/page.

\begin{acknowledgements}
This research has made use of the Astrophysics Data System, funded by NASA under Cooperative Agreement 80NSSC21M0056.
This research has made use of the SIMBAD database, operated at CDS, Strasbourg, France.
This work made use of data supplied by the UK Swift Science Data Centre at the University of Leicester.
We thank the anonymous referee for the constructive comments and insightful suggestions,
which helped to improve the clarity of the manuscript.
\end{acknowledgements}

\begin{appendix}

\section{Notes on individual analogous stars}
\label{sec:notes}

We collect here some useful notes on the properties of individual O- or B-type supergiants used for comparison with HMXB donor stars.

HD~209975 was classified as O9.5Ib by \citet{Prinja1990}, but most recent spectroscopic observations
reclassifies it as an O9Ib star\citep{deburgos2024a}, identification which is adopted here.

HD~92850 was classified as an O9.5I star by \citet{Prinja1990}, but \citet{deburgos2024a} identifies it as a B0Ia star,
identification which is retained here.

HD~204172 was classified as a B0Ib \citet{Prinja1990} and as a B0.5Ib star by \citet{Massa2024},
but we retain here the classification as B0.2Ia reported by \citet{deburgos2024a}.

HD~178487 and HD~148422, not present in de Burgos' catalogues,
have been both classified as B0.5Ib stars by \citet{Smartt1997}, and we retain here this identification.
However, we note that they were classified as B0Ia (HD~178487) and  B1Ia or B0.5Ia (HD~148422) by \citet{Prinja1990}.

HD~167264 is an O9.7Iab star \citep{Prinja1990, Sota2014, deburgos2024a}, a classification we  adopt here,
although it was also classified as a B0.5Iab star by \citet{Martins2015}.
We note that this star is present both in the list of O-type stars and among B-type stars
reported by \citet{Prinja1990}, probably due to a typographycal error.
We adopt here its wind properties reported by \citet{HP89}, \citet{Prinja1990} and \citet{Martins2015}.

HD~115842 is a B0.5Ia star according to SIMBAD \citep{MaizApellaniz2018}, but is    reported by \citet{Haucke2018} and \citet{deburgos2024a}
as a B0.5~Ia/ab star, therefore we do not include this star in our sample.

HD~91316 (aka $\rho$ Leo) has been classified as a B1Ib star in the literature,
but it is now accepted as a B1Iab star \citep{Negueruela2024}.

HD~13854 was classified as a B1Ia star \citep{Prinja1990}, but more recent papers report it as a B1Iab star \citep{Searle2008, deburgos2024a}, a classification we assume here.

HD~154090 is listed as a B1Ia star by \citet{Prinja1990}, a B0.7Ia by \citet{Crowther2006}, but we assume the most recent classification as a B2Iab star \citep{deburgos2024a}.

HD~47240 is a fast rotating B1Ib star and a pulsating variable, according to SIMBAD.
For this star very different wind terminal velocities have been reported, ranging from 450\,$\kms$ to 1050\,$\kms$.
In Table~C.1 we retain them all to estimate the average wind speed in the specific spectral type and LC.
However, we note that the lowest terminal velocity of 450\,$\kms$   \citep{Haucke2018} was measured from a spectrum where the H$_{\alpha}$ emission line was double-peaked,
indicative of a disk-like structure.
Moreover, we note a typographical error in \citet{Prinja1990}, where the object is listed as HD~47420 (B1Ib) instead of HD~47240. This appears to be a typo, as HD~47420 is a K4III star.

HD~52382 is reported in \citet{Prinja1990} as a B1Ia star, but the most recent classification is B0.5Ia \citep{deburgos2024a}, which we retain here.

HD~91943 is listed as a B0.5Ib star by \citet{Prinja1990}, as a B0.7Ia star by \citet{Crowther2006}, but we assume the most recent classification as a B0.7Ib star \citep{deburgos2024a}.

HD~163522 is reported  as a B1Iap star by \citet{Prinja1990}, and as a B1Ia by \citet{deburgos2024a} (and SIMBAD).

HD~99953 was classified as a B1.5Ia star \citep{Prinja1990}, but we assume the classification as a B2Ia star reported by \citet{deburgos2024a}.

HD~116084 is reported  as a B2.5Ib star \citep{Prinja1990}, but we assume the classification as a B2Ib star reported by \citet{deburgos2024a}.

HD~96248 was classified as a BC1.5Iab star \citep{Prinja1990}, but we assume the most recent classification as a B1Iab star reported by \citet{deburgos2024a}.

HD~148379 was classified as a B1.5Iap star  \citep{Prinja1990}, but we adopt here the  classification as a B2Iab  reported by \citet{deburgos2024a}.

HD~30614 was listed as a O9.5Ia star in   \citet{Prinja1990} and \citet{Markova2004},
but we assume the classification as a O9Ia star reported by \citet{deburgos2024a}.

HD~225160 is a O8Ib star in \citet{Markova2004}, but we retain the more recent classification as an O8Iab star \citet{Sota2014}.
We note that for HD\,225160 \citet{Markova2004} estimates the terminal velocity by
line-profile fitting of UV lines of IUE data, unlike other stars reported in this paper (as discussed previously).

HD~207198 is classified as a O9Ib/II star, not simply as an O9Ib star (e.g. \citealt{Markova2004}).

\onecolumn
\begin{landscape}
 \small
\section{Sample of HMXBs (SFXTs and classical SgXBs) with identified optical counterparts.}
\begin{longtable}{lrrlllrrrr}
\caption{Sample of HMXBs (SFXTs and SgXBs) with  identified optical counterparts and directly measured wind properties, when available (References are: Bu26=\citealt{Bulgarelli2026} Ch16=\citealt{Chaty2016}, C13=\citealt{Coleiro2013}, Coe96=\citealt{Coe1996},  Cox05=\citealt{Cox2005}, G12=\citealt{Goldoni2012}, GG14=\citealt{Gonzalez2014}, H20=\citealt{Hainich2020}, Ka06=\citealt{Kaper2006}, K21=\citealt{Kretschmar2021}, Mas09=\citealt{Mason2009}, Ma15=\citealt{Martinez2015},  MaAp18=\citealt{MaizApellaniz2018}, Ne08=\citealt{Nespoli2008}, Ne10=\citealt{Nespoli2010}, N06=\citealt{Negueruela2006},  NS7=\citealt{Negueruela2007}, N08=\citealt{Negueruela2008}, N19=\citealt{Negueruela2019}, Pe06=\citealt{Pellizza2006},  Re96=\citealt{Reig1996}, Ro10=\citealt{Romano2010igr18483}, Ro15=\citealt{Romano2015}, Ro23=\citealt{Romano2023}, So14=\citealt{Sota2014}, SP18=\citealt{Sidoli2018}, Si23=\citealt{Sidoli2023xte1739}, To10=\citealt{Torrejon2010}).
} \\
\hline\hline
\label{tab:list1}
X-ray source &       DR &  Ref  &  Opt. count &  Sp.type & LC      &  Ref & v$_{\infty}$/10$^3$  ($\kms$)  & $\mdot_w$ ($\msunyr$)    & Ref  \\
\hline
\endfirsthead
\caption{continued.}\\
\hline\hline
X-ray source &       DR &  Ref  &  Opt. count &  Sp.type & LC      &  Ref & v$_{\infty}$/10$^3$ & $\mdot_w$    & Ref  \\
             &           &       &     &          &         &      &  ($\kms$)           & ($\msunyr$)   &      \\
\hline
\endhead
\hline
\endfoot
Estreme SFXTs         &                    &       &            &         &          &     &                     &              &        \\

\hline
IGRJ17544-2619 & 1.7$\times10^{6}$ &  SP18  &             & O9     & Ib  & Pe06   &   1.5$^{+0.2} _{-0.2}$ &  1.6$^{+0.9} _{-0.6}\times10^{-6}$   &  GG16   \\
XTEJ1739-302   &  3$\times10^{5}$  &  Si23  &             & O8     & Iab  & N06   &     &     &     \\
IGRJ16479-4514 &  16500           &  Ro23   &            & O8.5      & Ib  & N19  &     &     &     \\
               &                  &         &            & O9.5      & Iab & Ne08  &     &     &     \\
IGRJ08408-4503 &  15350            & Ro23   &  LM Vel    & O8.5      & Ib-II(f)p   & So14, N08  &   1.9$^{+0.1} _{-0.1}$   &  7.9$^{+4.6} _{-2.9}\times10^{-7}$   &  H20 \\
AXJ1841.0-0536 &  1.1$\times10^{4}$ &   SP18 &           & B0.2     & Ibp  & N08, N19   &     &     &     \\
SAXJ1818.6-1703&   8600           &  Ro23    &            & B0.5      & Iab  & To10  &     &     &     \\
\hline
Other SFXTs         &                    &       &            &         &          &     &                     &              &        \\
\hline
IGRJ11215-5952 &   $>$480         &  SP18     & HD 306414    & B0.5      & Ia   & N19  &   0.8$^{+0.2} _{-0.1}$   &  3.16$^{+1.8} _{-1.2}\times10^{-7}$   &  H20 \\
IGRJ00370+6122 &    300           & this paper & BD+60 73    & BN0.7     & Ib   & GG14  &  1.1$^{+0.1} _{-0.2}$   &  3.16$^{+0.8} _{-1.2}\times10^{-8}$   &  H20   \\
IGRJ16465-4507 &    40            &  SP18     &             & B0.5      & Ibn  & N19        &     &     &     \\
               &                    &    -    &             & B0.5-B1   & Ib    & G13, Ch16  &     &     &     \\
IGRJ18483-0311 & 1200              &    Ro10   &             & B0.5-B1     & Iab  & To10  &     &     &     \\
IGRJ17354-3255 &   $>$2500          &   Bu26   &             & O8.5-O9    & Iab  & C13  &     &     &     \\
IGRJ16418-4532 &    660            &  Ro23    &             & BN0.5     & Ia  & C13  &     &     &     \\
               &                   &          &              & O9.5      & I   & G12  &     &     &     \\
IGRJ16328-4726 &   300             &   SP18   &              & O8        & Iaf (or Iafpe) &  C13  &     &     &     \\
AXJ1845.0-0433 &   1600            &  Ro23     &            & O9     & Ia  & N08   &     &     &     \\
               &                   &           &            & O9.5   & I  & Coe96   &     &     &     \\
IGRJ16195-4945 &    100           & this paper &            & ON9.7  & Iab & C13   &     &     &     \\
\hline
Classical SgXBs &                &          &              &         &          &     &                     &              &        \\
\hline
Vela X-1     &   1.7             & SP18     & HD 77581     & B0.5 & Ia   & MaAp18   &  0.7$^{+0.2} _{-0.1}$    &  6.3$^{+3.7} _{-2.3}\times10^{-7}$    &  GG16   \\
4U1700-37    &   12             & SP18      &   HD153919   & O6      & Iafcp   & So14  &   1.9$^{+0.1} _{-0.1}$   &  2.5$^{+1.5} _{-1.2}\times10^{-6}$   &  H20 \\
4U1538-52     &   33           &  SP18      &             &  B0      & I  &  N19   &     &     &     \\
4U1909+07     &   12            & SP18      &             &  B0-B3   & I  &  Ma15   & 0.5$^{+0.1} _{-0.1}$     &  2.8$\times10^{-7}$    &  Ma15   \\
SAXJ1802.7-2017 &   375        &  SP18      &             &  B1      & Ib  &  N19   &     &     &     \\
XTEJ1855-026   &     1         &    SP18    &             & BN0.2    & Ia  & N19   &     &     &     \\
IGRJ16493-4348 &     50        &  Ro23      &             & B0.5-B1  & Ia-Ib  & Ne10   &     &     &     \\
4U1907+097     &    550    & SP18    &             &  O8-O9   & I  &  Cox05   &     &     &     \\
IGRJ16320-4751 &   15     &  SP18    &            & BN0.5    & Ia   & C13   &     &     &     \\
EXO1722-363    &   17     &  SP18    &            & B0-B1    & Ia   & Mas09   &     &     &     \\
OAO1657-415    &   10     &  SP18    &            & O/WN9    & fpe   & Mas09   &     &     &     \\
2S0114+65      &   1     &   SP18    &           & B1   & Ia   & Re96   &     &     &     \\
IGRJ19140+0951 &   770     &  SP18   &             & B0.5    & Ia  & To10   &     &     &     \\
1E1145.1-6141  &   1      &  SP18    &            & B2    & Ia  & N19   &     &     &     \\
GX301-2        &   2.6     & SP18    &             & B1    & Ia+  & N19   &  0.305  & 1.0$\times10^{-5}$    &  Ka06   \\
IGRJ16207-5129 &   9     &  SP18     &           & B1      & Ia    & Ne08   &     &     &     \\
               &         &       &          & B0      & I     & NS07   &     &     &     \\
\hline
\end{longtable}
\end{landscape}
\FloatBarrier 
\clearpage

\onecolumn
\begin{landscape}
  \small
\section{Stellar analogs of optical counterparts in our sample of HMXBs}
\begin{longtable}{lllrrrccccr}
\caption{Stellar analogs of optical counterparts in our sample of HMXBs, with published wind terminal velocities and/or  mass-loss rates.
Refs are: Be01=\citealt{Benaglia2001},  Cr06=\citealt{Crowther2006}, dB24=\citealt{deburgos2024a}, Ha18=\citealt{Haucke2018}, M07=\citealt{Mokiem2007a}, Ma04=\citealt{Markova2004}, Ma24=\citealt{Massa2024}, Mar15=\citealt{Martins2015}, MaPu08=\citealt{Markova2008}, Hol25=\citealt{holgado2025}, HP89=\citealt{HP89}, P90=\citealt{Prinja1990}, PH86=\citealt{PH86} Se08=\citealt{Searle2008}, S97=\citealt{Smartt1997}.
} \\
\hline
\hline
\label{tab:analogs}
Sp.type & LC  &  Name    &   $v_{\infty}$ &  $\mdot_w$    &  Refs  &  $T_{eff}$  &   $R_{\star}$    &    $M_{\star}$   &   Log($L_{\star}$)  &  Refs \\
        &     &          &     ($\kms$)   & ($\msunyr$)   &        &    (kK)     &   ($R_\odot$)    &    ($M_\odot$)   &    ($L_\odot$)      &      \\
\hline
\endfirsthead
\caption{continued.}\\
\hline\hline
Sp.type & LC  &  Name    &   v$_{\infty}$ &  $\mdot_w$    &  Refs  &  T$_{eff}$  &   $R_{\star}$    &    $M_{\star}$   &   Log($L_{\star}$)  &  Refs \\
        &     &          &     ($\kms$)   & ($\msunyr$)   &        &    (kK)     &   ($R_\odot$)    &    ($M_\odot$)   &    ($L_\odot$)      &      \\
\hline
\endhead
\hline
\endfoot
O8       & Ia & HD151804 &  1445 &  7.94$\times$10$^{-6}$ & P90, HP89    &    -    &    -             &   -              &           -          &  -  \\
O8       & Iab& HD225160 &  1600 &  5.3$\times$10$^{-6}$  & Ma04  &   33.0  &   22.4          &   -              &   5.73         &  Ma04  \\
O8       & Ib & HD167971 &  2185 &  3.16$\times$10$^{-6}$ & P90,HP89  &    -    &    -             &   -              &           -          &  -  \\
\hline
O8.5     & Ia &  -        & -     &   -                    &  -        &    -    &    -             &   -              &           -          &  -  \\
O8.5     & Iab & HD112244    & 1575 & 2.51$\times$10$^{-6}$ & P90,HP89  &     -    &    -             &   -              &           -          &  -  \\
         &   & HD112244    & 1880 &                        &  Be01  &     -    &    -             &   -              &           -          &  -  \\
         &   & HD112244    & 1950 &                        &  Be01  &     -    &    -             &   -              &           -          &  -  \\
         &   & HD112244    & 2160 &                        &  Be01  &     -    &    -             &   -              &           -          &  -  \\ 
O8.5     & Ib & HD96917      & 2000 & 2.00$\times$10$^{-6}$ & P90,HP89  &     32.0  &    21.1     &  29            &  5.62            &      Hol25 \\
\hline
O9       & Ia & HD149404 &  2450 &  3.16$\times$10$^{-6}$ & P90,HP89  &    -    &    -             &   -              &           -          &  -  \\
O9       & Ia & HD30614  & 1590   & 7.94$\times$10$^{-7}$ &  P90,HP89 &    29.0    &    -             &   -              &           -          &  dB24  \\
         &    & HD30614  & 1900  &                        &  PH86     &            &    -             &   -              &           -          &     \\
O9       & Iab & HD210809  &  2135 &  1.26$\times$10$^{-6}$ & P90,HP89   &   31.0  &   22.0          &   39             &     5.6          &  HP89  \\
         &     & HD210809  &  2100 &  5.3$\times$10$^{-6}$  & M07     &   32.7  &   -          &   -             &     -         &  dB24 \\
         &     & HD210809  &  -    &  -     & -     &   31.5  &   21.2        &   -             &     5.6         &   M07  \\
O9       & Ib & HD207198  &  2090 &  3.98$\times$10$^{-7}$ & P90,HP89  &  33.0   &    16.6          &   -              &        5.47          &  M07   \\
         &    & HD207198  &  2150 &  1.76$\times$10$^{-6}$ & M07       &    -    &    -             &   -              &           -          &  -  \\
         &    & HD207198  &  2100 &  -                     &  Ma24     &  32.0    &    -            &   -              &           -          &  Ma24  \\
         &    & HD207198  &  2100 &  9.0$\times$10$^{-7}$  & Ma04 &  31.7  &   15.2          &   -              &   5.32          &  Ma04  \\
O9       & Ib & HD209975  &  2000 &  3.16$\times$10$^{-7}$ &  Mar15    &  30.5    &   17            &   -              &       5.35          &  Mar15  \\
         &    & HD209975  &  2010  & 5.01$\times$10$^{-7}$ & P90,HP89  &  33.9    &                 &                  &                     &  dB24  \\
         &    & HD209975  &  2050 & 1.8$\times$10$^{-6}$ &  Ma04  &  31.0    &  20.9          &   -              &  5.56          &  Ma04  \\
         &    & HD209975  &       & 1.58$\times$10$^{-6}$ & Ma04  &          &  19.2          &                  &  5.49          &  Ma04  \\
\hline
O9.5     & Ia  &  none &    &    &    &      &             &                &              &    \\
O9.5     & Iab & HD154368  &  1850 &                        & P90          &  31.5     &             &                &              & dB24   \\
O9.5     & Iab & HD188209  &  1650 &                        & P90          &  31.6     &             &                &              & dB24   \\
O9.5     & Iab & HD218915  &  1830 &                        &  P90         &   -       &             &                &              &    \\
O9.5     & Ib  &  none     &       &                       &         &           &             &                &              &    \\
\hline
O9.7     & Ia  & none  &    &    &    &      &             &                &              &    \\
O9.7     & Iab & HD167264  &  1795  &  6.3$\times$10$^{-7}$  & P90, HP89                 & 28.2 &  21 &  33  &   5.5    &  dB24, HP89   \\
         &     & HD167264  &  2000  &  3.16$\times$10$^{-7}$ & Mar15,Mar15 & 28.0 &  29 &      &   5.65   &  Mar15   \\
O9.7     & Iab &  HD75222 &  1840   &  6.3$\times$10$^{-7}$  &  P90,HP89   &  28.1    &   22.2      &    26.7        &    5.53      &  dB24, Hol25    \\
O9.7     & Iab & HD149038   & 1830  &  6.3$\times$10$^{-7}$   & P90,HP89     &  28.3    &             &                &              &  dB24 \\
         &     & HD149038   & 2200  &  1.96$\times$10$^{-6}$  & PH86,PH86    &  30.0    &             &                &              &  PH86 \\
O9.7     & Ib  &  HD47432  &  1590  &                       &   P90                     &  27.8    &  21.4         &   19.4   &    5.49    &  dB24, Hol25  \\
         &     &  HD47432  &  1600  &  1.9$\times$10$^{-6}$  &  Ma04, Ma04 & 30.5     &  18.9         &          &    5.45    &  Ma04  \\
\hline
\hline
B0      & Ia & HD37128  & 1910   &  2.68$\times$10$^{-6}$  &  P90, PH86   &  25.0     &     34        &    37            &               &  PH86  \\
B0      & Ia  & HD37128  &        &  2.25$\times$10$^{-6}$  &  M07       &  27.0      &     24        &                  &    5.44       &  M07, Cr06  \\
        &   & HD37128  & 1600   &  2.00$\times$10$^{-6}$   & Se08      &  27.5      &    32.4       &     40           &   5.73        & Se08   \\
        &   & HD37128  & 1800   &  5.62$\times$10$^{-7}$  & Mar15      &  27.5      &    28          &    -            &    5.6        &  Mar15   \\
         &   & HD37128  &        &                         &            &  27.1      &               &                &               &  dB24  \\
         &   & HD37128  & 1700   &                         &  Ma24 &  28.0      &               &                &               &  Ma24   \\
         &   & HD37128  & 1980   &  2.68$\times$10$^{-6}$  &  PH86,PH86 &  25.0      &               &                &               &  PH86   \\
B0       & Ia & HD91452 & 1635   &                         &  P90       &  -          &       -       &   -             &          -    &   - \\
B0       & Ia & HD94909 & 1050   &  2.00$\times$10$^{-6}$  &  P90, M07, Cr06 &  27.0   &     25.5      &   -             &     5.49     &   M07, Cr06  \\
B0       & Ia & HD92850  & 1615  &   6.31$\times$10$^{-7}$ & P90, HP89 &  30.2   &   20      &   34   &     5.5    &      HP89   \\
         &    & HD92850  &       &                         &           &  28.3   &          &      &             &    dB24 \\
B0       & Ia & HD122879  & 1620 &   3$\times$10$^{-6}$   &  M07, Cr06 &  28.0    &   24.4          &    -            &   5.52           &    M07, Cr06  \\
         &    & HD122879  &   -  &                        &            &  27.4    &    -           &     -           &                   &  dB24   \\
B0       & Ib &  HD164402 &  1650  &                       &  P90       & 28.9     &             &                &              & dB24   \\
         &    &  HD164402 &  1680  &                       &  Ma24 & 28.0     &             &                &              & Ma24  \\ 
B0       & Ib &  HD167402 &  2005  &                       &  P90       &          &             &                &              &    \\
B0       & Ib &  HD192660  & 1850  &  5$\times$10$^{-6}$   &  Se08, Se08  &  30      &    23.4          &    33            &     5.74         & Se08   \\ 
\hline
B0.2 &  Ia   &    none      &      &                         &           &              &             &                &               &     \\
B0.2 &  Iab  & HD204172     & 1685  & 5.7$\times$10$^{-7}$   &  Se08, Se08  &   28.5       &    22.4     &    41          &     5.48     &  Se08  \\
     &       & HD204172     & 1630  &                        &  P90                     &   27.3       &             &                &              &   dB24      \\
     &       & HD204172     & 1800  &                        &  Ma24               &              &             &                &              &             \\
B0.2 &  Ib   & HD113012     & 1215  &                        &  P90      &              &             &              &               &        \\
\hline
B0.5 &   Ia     & HD38771      &  1500 &   1.4$\times$10$^{-7}$ & Ha18             &   25.0      &    13     &                &    4.78     &  Ha18   \\
     &          & HD38771      &  1525 &                        &  P90                   &   26.0      &    28     &   34           &             &  PH86   \\
     &          & HD38771      &  1390 &  1.2$\times$10$^{-6}$  & Se08, Se08 &             &           &                &    5.48         &  Se08   \\
     &          & HD38771      &  1800 &  1.96$\times$10$^{-6}$ &     HP86, HP86         &   26.7      &           &                &                &  dB24  \\
     &          & HD38771      &  1525 &   9.0$\times$10$^{-7}$ &   Cr06         &   26.5      &    22.2   &                &    5.35        &    Cr06, M07   \\
B0.5 &   Ia     & HD52382       &  900 &                        &     P90                &   23.6      &          &                 &               &   dB24    \\
B0.5 & Ia (N wk) & HD152234    & 1450 & 2.7$\times$10$^{-6}$ & Cr06, Cr06 &   26.0     & 42.4      &                &   5.87       &  Cr06   \\
     &           & HD152234    &       &                      &                           &   26.1     &           &                &              &   dB24   \\
B0.5 & Ia       & HD152235     &   850 &                        &  P90                  &   24.0       &            &                &                &  dB24   \\
B0.5 & Iab      &  none        &      &                        &                         &            &             &                 &              &     \\
B0.5 &  Ib      & HD155985    &   1225   &                       &   P90                  &  25.1      &            &                &             &  dB24   \\
B0.5 &  Ib      & HD213087    &   1520   &  7.0$\times$10$^{-7}$ &  P90, Se08       &  27.0      &   32       &   40           &  5.69      &  Se08   \\
     &          & HD213087    &          &                       &                        &  25.6      &            &                &            &  dB24  \\
B0.5 &  Ib      &  HD178487   &    1520  &                         &  P90                    &  27.0      &             &                &          &  S97   \\
B0.5 &  Ib      &  HD148422   &    1335  &                         & P90                     &  25.0      &             &                &          &  S97   \\
B0.5 &  Ib      &  HD64760    &    1600  &  1.1$\times$10$^{-6}$   & Se08, Se08  &  28.0      &   23.3    &     33        &    5.48  &  Se08   \\
     &          &  HD64760    &    1650  &                         & Ma24               &  25.0      &           &               &          &  Ma24, dB24   \\
     &          &  HD64760    &    1500  &  4.2$\times$10$^{-7}$   & Ha18, Ha18  &  23.0      &  12       &               &   4.57   &  Ha18 \\
\hline
B0.7 &  Ia     &   HD2905    &   1105    &   2.0$\times$10$^{-6}$  & P90, Cr06      &   24.4           &             &                &               & dB24    \\
     &         &   HD2905    &    850    &   2.5$\times$10$^{-6}$  & Se08, Se08 &   23.5           &    33       &                 &   5.48      & Se08    \\
     &         &   HD2905    &           &                         &                        &   21.5           &   41.4      &                 &   5.52       &  Cr06     \\
B0.7 &  Iab     &  none        &       &                      &                      &              &             &                &               &     \\
B0.7 &  Ib      &  HD91943     &  1405  &                      &  P90                   &      26.7        &             &                &               &   dB24     \\
     &          &  HD91943     &  1470  &  7.5$\times$10$^{-7}$ &  M07, Cr06     &     24.5         & 26.3        &           &    5.35   &   M07, Cr06     \\
B0.7 &  Ib      &  HD109867    & 1155    &                       &   P90                 &              &             &                &               &     \\
B0.7 &  Ib      &  HD190066    & 1275    &   7$\times$10$^{-7}$  & Se08, Se08 &   21.0      &     41.4        &   33             &    5.54   & Se08    \\
     &          &  HD190066    &         &                       &                        &   24.8      &                 &                  &           & dB24     \\
\hline
B1  &  Ia       &  HD 40111    &   1465   &                      &  P90                   &              &             &            &               &     \\
B1  &  Ia       &  HD148688    &    725   & 1.75$\times$10$^{-6}$ &  P90, Cr06    &    22.0       &   36.7      &            &    5.45      &  Cr06   \\
B1  &  Ia       &  HD160993    &   1320   &                      & P90                   &               &             &             &             &          \\
B1  &  Ia       &  HD150168    &   1385   &                      & Ma24     &    25.0           &             &             &             &  Ma24     \\
B1  &  Ia       & HD163522     &   1240   &                      & P90       &   24.2            &             &             &             &    dB24  \\
B1  &  Iab      & HD29138 & 1315     &                        &    P90            &          &             &             &             &    \\
B1  &  Iab      & HD58510  &  930    &                        &   P90               &          &             &             &             &    \\
B1  &  Iab      & HD91316  &  1110    & 3.5$\times$10$^{-7}$  &   P90, Cr06  &    22.0      &  37.4       &             &  5.47       & Cr06   \\
    &           & HD91316  &  1300    &                       &   Ma24          &    25.0      &             &             &             &  Ma24   \\
    &           & HD91316  &  1350    &  1.31$\times$10$^{-6}$ &   PH86, PH86         &   21.0      &             &             &             &     \\
B1  &  Iab      & HD13854    &  920    & 8.5$\times$10$^{-7}$ &  P90, Cr06      &   20.0          &   49.2      &     33    &   5.54        & Se08    \\
    &           & HD13854    &  955    & 1.5$\times$10$^{-6}$ &  Se08, Se08 &   22.5          &             &           &               &  dB24   \\
B1  &  Iab      & HD96248    & 675     &                      & P90             &    22.1       &             &           &               &  dB24   \\
B1  &  Ib       & HD24398    &  1270   &                      & P90                &     23.5      &             &             &             &  dB24   \\
    &           & HD24398    &  1200   &                      & Ma24              &     23.0      &             &             &             & Ma24    \\
B1  &  Ib       & HD47240    &  960    &                      &      P90          &   21.9        &         &             &             &  dB24   \\
    &           & HD47240    & 1050    &                      &  Ma24             &   24.0        &             &             &             &  Ma24   \\
    &           & HD47240    &  450    & 2.4$\times$10$^{-7}$ & Ha18, Ha18 &    19.0       &    30       &             &   5.03      & Ha18    \\
B1  &  Ib       & HD86606   &    490   &                      &  P90                   &               &             &             &             &     \\
B1  &  Ib       & HD100276  &  1430    &                      &   P90                  &               &             &             &             &     \\
B1  &  Ib       & HD104683  &   1145   &                      &    P90                 &               &             &             &             &     \\
B1  &  Ib       & HD119608  &  880    &                       &    P90                &     22.0      &             &             &             &  dB24  \\
B1  &  Ib       & HD157246  &   735  &                       &    P90                 &      23.7     &             &             &             &  dB24  \\
    &           & HD157246  &  1000  & 3.65$\times$10$^{-7}$ &   PH86,PH86            &      21.0     &             &             &             &  PH86  \\
B1  &  Ib       & HD191877 &   1160  &                        &    P90                &      23.6     &             &             &             &  dB24  \\
B1  &  Ib       & HD235783 &   1070  &                        &    P90                &     22.9      &             &             &             &  dB24    \\
B1  &  Ib       & HD179407 &   1585  &                        &    P90                &     24.5      &             &             &             &   S97    \\
B1  &  Ib       & HD99857 &  1705    &                        &    P90                &     24.4     &             &             &             &   dB24  \\
\hline
B2   &  Ia       &  HD14143 &   645  & 1.05$\times$10$^{-6}$ &  P90, Cr06     &    18.0     &   52.9       &            &  5.42       &   Cr06     \\
     &           &  HD14143 &       &                        &                        &    19.9    &              &            &              &   dB24  \\
B2   &  Ia       &  HD14818 &  565   &  5.5$\times$10$^{-7}$ &   P90, Cr06    &   18.5      &   46.1     &              &     5.35     &   Cr06     \\
     &           &  HD14818 &  625   & 1.0$\times$10$^{-6}$ & Se08, Se08  &    18.0     &   51.4     &   26         &     5.4      &    Se08  \\
     &           &  HD14818 &        &                      &                         &    20.3     &            &               &              &     dB24    \\
B2   &  Ia       &  HD41117 & 510   & 9.0$\times$10$^{-7}$  &    P90, Cr06  &    19.0     &   61.9     &            &     5.65          & Cr06     \\
     &           &  HD41117 &       &                       &                       &    19.8     &            &            &                   & dB24 \\
     &           &  HD41117 & 510   & 1.7$\times$10$^{-7}$  & Ha18, Ha18 &  19.0      &   23       &            &    4.84           &  Ha18 \\
B2   &  Ia       &  HD99953 & 510  &                         &     P90                &     20.3   &            &            &                   & dB24 \\
B2   &  Ia       & HD194279 & 550   &  1.05$\times$10$^{-6}$ & Cr06, Cr06 & 19.0   &   44.7          &            &   5.37      &   Cr06   \\
     &           & HD194279 &       &                        &                        & 19.4         &           &           &             &  dB24    \\
B2  &  Iab       &  HD154090    &   915   &                &  P90            &  23.4          &             &            &               &  dB24   \\
B2  &  Iab       &  HD148379    &   510   &                &  P90            &  18.5          &             &            &               &  dB24   \\
B2   &  Ib       &  HD93840 & 1160  &                        &   P90             &              &             &            &               &      \\
B2   &  Ib       &  HD97707 & 870   &                        &   P90             &              &             &            &               &      \\
B2   &  Ib       &  HD116084 & 550 &                         &   P90             &    19.4      &             &            &               &  dB24    \\
B2   &  Ib       &  HD165024 & 1130 &                        &   P90             &    22.5      &             &            &               &  dB24    \\
B2   &  Ib       &  HD206165 & 640  & 5.0$\times$10$^{-7}$   &  P90, Se08  &    18.0      &   39.8      &    21      &   5.18       & Se08     \\
     &           &  HD206165 &      &                        &                   &    19.3      &   32        &    12      &   5.11       &   MaPu08   \\
     &           &  HD206165 &      &                        &                   &    19.8      &             &            &              &  dB24  \\
\hline
B3   &  Ia      &  HD53138     &  830 &                      &  P90                    &   16.5       &   54.7      &     23     &    5.3        & Se08 \\
     &          &  HD53138     &  600 & 2.4$\times$10$^{-7}$ &  Ha18, Ha18 &   16.8       &             &            &               &  dB24    \\
     &          &  HD53138     &  450 & 2.0$\times$10$^{-7}$ &  Ha18, Ha18 &   18.0       &   46        &            &    5.31       &  Ha18    \\
     &          &  HD53138     &  500 & 4.5$\times$10$^{-7}$ &  Se08, Se08 &              &             &            &               &      \\
B3   &  Ia      &  HD198478    & 470  &                &    P90           &              &             &            &               &      \\
B3   &  Ia       &  HD14134   &  465 & 5.2$\times$10$^{-7}$  & Cr06, Cr06  &   16.0   &  56.7     &            &   5.28         & Cr06     \\
B3   &  Ia       &  HD75149 & 400 &  0.9$\times$10$^{-7}$ & Ha18, Ha18  &    16.0      &    61        &               &    &    Ha18 \\
    &          &  HD75149  & 350 &  2.0$\times$10$^{-7}$ & Ha18, Ha18  &              &             &            &     &      \\
    &          &  HD75149 & 400 &  1.6$\times$10$^{-7}$ & Ha18, Ha18  &              &             &            &      &      \\
    &          &  HD75149 & 350 &  2.5$\times$10$^{-7}$ & Ha18, Ha18  &              &             &            &    &      \\
B3  &  Ia       &  HD53138     &  865     & 3.6$\times$10$^{-7}$  &   Cr06, Cr06 &   15.5    &  65     &         &   5.34   & Cr06     \\
B3  &  Iab        &  HD43384     &  710   &            &      P90         &    15.9          &             &            &               & dB24     \\
B3  &  Ib        &  none    &       &                &               &             &             &            &               &     \\
\hline
\end{longtable}
\end{landscape}

\FloatBarrier

\begin{table*}
  \small
\caption{Average wind properties of the OB supergiants listed in Table~\ref{tab:analogs}, for each spectral type and luminosity class (LC).
We report the mean and the covered range of values (the range is ``$-$'' if only one value is listed in Table~\ref{tab:analogs}).
For B1Ib stars we report two values for the terminal wind velocity, one excluding the   450~$\kms$ measurement for HD~47240
(obtained during a phase with a disk-like wind, Sect.~\ref{sec:notes}), and a second value, given in parentheses, which includes this lowest velocity.
}
\label{tab:average}
\centering
\begin{tabular}{llrccc}
\hline\hline
Sp.type & LC &  $\overline{v}_\infty$ [km\,s$^{-1}$] &    ${v}_\infty$ range    &  $\overline{\dot{M}}$ [$M_\odot\,\mathrm{yr^{-1}}$]  & $\dot{M}$ range \\
\hline
O8 & Ia     & 1445         &    $-$      &  7.9$\times$10$^{-6}$  &  $-$    \\
O8 & Iab    & 1600         &    $-$      &  5.3$\times$10$^{-6}$  &   $-$         \\
O8 & Ib     & 2185         &    $-$      &  3.16$\times$10$^{-6}$ &   $-$  \\
\hline
O8.5 & Ia   & none         &   none      & none                   &  none  \\ 
O8.5 & Iab  & 1890         & 1575--2160  &  2.51$\times$10$^{-6}$  &   $-$  \\
O8.5 & Ib   & 2000         &    $-$      &  2.0$\times$10$^{-6}$   &   $-$  \\
\hline
O9   & Ia  & 1870          & 1550--2450   & 2.7$\times$10$^{-6}$   &   (0.80--4.2)$\times$10$^{-6}$   \\
O9   & Iab & 2040          & 1820--2135   & 3.0$\times$10$^{-6}$   &   (0.80--5.3)$\times$10$^{-6}$   \\
O9   & Ib  & 2070          & 2050--2150   & 1.0$\times$10$^{-6}$   &   (0.32--1.8)$\times$10$^{-6}$ \\
\hline
O9.5 & Ia   &  none          &  none        &  none                 &   none   \\
O9.5 & Iab  &  1780          &  1650--1850  &  none                 &   none   \\
O9.5 & Ib  &   none          &  none        & none                  &  none   \\
\hline
O9.7 & Ia   &  none        &   none         & none                  &    none  \\
O9.7 & Iab  &  1930        &  1795--2200    & 8.3$\times$10$^{-7}$  &  (3.2--19.6)$\times$10$^{-7}$   \\
O9.7 & Ib   &  1590        &  1590--1600    & 1.9$\times$10$^{-6}$  &  none   \\
\hline
\hline
B0   & Ia   & 1660         &  1050--1980    & 2.0$\times$10$^{-6}$  &  (0.6--3.0)$\times$10$^{-6}$   \\
B0   & Iab  & none         &   none         & none                  &    none       \\
B0   & Ib   & 1800         &  1650--2005    &  5$\times$10$^{-6}$   &   $-$      \\
\hline
B0.2 & Ia  &  none          &  none         & none                 & none   \\
B0.2 & Iab & 1710          &   1630--1800   & 5.7$\times$10$^{-7}$  &  $-$      \\
B0.2 & Ib  & 1215          &    $-$         & none                 & none \\
\hline
B0.5 & Ia   & 1390         &   850--1800   & 1.4$\times$10$^{-6}$ &  (0.14--2.7)$\times$10$^{-6}$   \\
B0.5 & Iab  & none          & none   & none          & none  \\
B0.5 & Ib   & 1480          &  1225--1650   & 7.4$\times$10$^{-7}$   &   (4.2--11.0)$\times$10$^{-7}$   \\
\hline
B0.7 & Ia   & 980           &   850--1105   & 2.25$\times$10$^{-6}$   &   (2.0--2.5)$\times$10$^{-6}$  \\
B0.7 & Iab  & none          & none          & none                   & none    \\
B0.7 & Ib   & 1330          & 1155--1470    & 7.25$\times$10$^{-7}$   &  (7.0--7.5)$\times$10$^{-7}$   \\
\hline
B1 & Ia   & 1230           &   725--1465    & 1.75$\times$10$^{-6}$    &   $-$    \\
B1 & Iab  & 1070           &   675--1350    & 1.0$\times$10$^{-6}$  &   (0.35--1.5)$\times$10$^{-6}$  \\
B1 & Ib   & 1120 (1075)    &   735--1705 or (450)--1705    & 3.0$\times$10$^{-7}$  &   (2.4--3.65)$\times$10$^{-7}$ \\
\hline
B2 & Ia   & 560            &   510--645    &  7.9$\times$10$^{-7}$   &  (1.7--10.5)$\times$10$^{-7}$  \\
B2 & Iab  & 710            &   510--915    &   none                  & none \\
B2 & Ib   & 870            &   550--1160   &  5.0$\times$10$^{-7}$    &  $-$      \\
\hline
B3 & Ia   & 516           &   350--830    &  2.7$\times$10$^{-7}$    &  (0.9--5.2)$\times$10$^{-7}$  \\
B3 & Iab  & 710           & $-$           &      none               &   none \\
B3 & Ib   &  none         & none          & none                    & none \\
\hline
\end{tabular}
\end{table*}

\clearpage

\section{Stellar analogs }
\scriptsize
\begin{longtable}{llllllll}
  \caption{Properties of stellar analogs selected exclusively from \citet{deburgos2024a}. }\\
\hline\hline
\label{tab:teffqlog}
Star  &       Sp.type  \& LC   & T$_{eff}$   &  Log(Q)      &  Star  &       Sp.type  \& LC   & T$_{eff}$   &  Log(Q)  \\
        &                        &  (K)        &           &          &                        &  (K)        &        \\
\hline
\endfirsthead
\caption{continued.}\\
\hline\hline
Star  &       Sp.type  \& LC   & T$_{eff}$   &  Log(Q)      &  Star  &       Sp.type  \& LC   & T$_{eff}$   &  Log(Q)  \\
        &                        &  (K)        &           &          &                        &  (K)        &        \\
\hline
\endhead
\hline
\endfoot
HD30614  &   O9Ia  &         29000    &         -12.5         &   HD163522  &   B1Ia  &         24200    &         -13.2    \\
HD202124  &   O9Iab  &         29100    &         -12.5     &   HD167287  &   B1Iab  &         26400    &         -13.7    \\
HD148546  &   O9Iab  &         30800    &         -12.5     &  HD194153  &   B1Iab  &         24900    &         -13.2    \\
HD210809  &   O9Iab  &         32700    &         -12.5     &   HD224424  &   B1Iab  &         22200    &         -13.1       \\
HD209975  &   O9Ib  &         33900    &         -12.6       &     HD15785  &   B1Iab  &         24100    &         -13.1     \\
BD+6312  &   O9Ib  &         31900    &         -13.9         &     HD96248  &   B1Iab  &         22100    &         -13.1     \\
HD151018  &   O9Ib  &         31600    &         -12.5      &      HD13854  &   B1Iab  &         22500    &         -13.3      \\
HD188209  &   O9.5Iab  &         31600    &         -12.5  &      HD159864  &   B1Ib  &         27300    &         -13.7      \\
HD237211  &   O9.5Iab  &         33600    &         -12.5  &      HD166287  &   B1Ib  &         24700    &         -14.0   \\
HD154368  &   O9.5Iab  &         31500    &         -12.5  &      HD166540  &   B1Ib  &         28000    &         -14.0    \\
HD60479  &   O9.5Ib   &         30700    &         -13.4     &     HD168489  &   B1Ib  &         27400    &         -13.9   \\
HD195592  &   O9.7Ia  &         27100    &         -12.5     &    HD166689  &   B1Ib  &         25000    &         -13.8      \\
HD167264  &   O9.7Iab  &         28200    &         -13.2  &    HD170604  &   B1Ib  &         26800    &         -13.9     \\
HD225146  &   O9.7Iab  &         27400    &         -13.3  &    HD169673  &   B1Ib  &         26600    &         -14.0    \\
HD75222  &   O9.7Iab  &         28100    &         -13.0    &    HD167451  &   B1Ib  &         26000    &         -13.5   \\
HD149038  &   O9.7Iab  &         28300    &       -13.2    &     HD191877  &   B1Ib  &         23600    &         -14.0   \\
HD165319  &   O9.7Ib  &         28700    &         -13.1    &   HD331759  &   B1Ib  &         27800    &         -13.9    \\
HD18409  &   O9.7Ib  &         29500    &           -13.5    &    HD194057  &   B1Ib  &         25800    &         -13.2  \\
HD47432  &   O9.7Ib  &         27800    &           -13.1    &    BD+483437  &   B1Ib  &         25000    &         -13.8   \\
HD164971  &   B0Ia  &         27500    &            -13.6    &   HD205139  &   B1Ib  &         25200    &         -13.6    \\
HD171201  &   B0Ia  &         31300    &            -13.1     &  HD235783  &   B1Ib  &         22900    &         -14.0      \\
HD168021  &   B0Ia  &         27300    &            -13.2     &   HD218941  &   B1Ib  &         23400    &         -13.7  \\
HD37128   &   B0Ia  &         27100    &             -13.1     &    BD+612529  &   B1Ib  &         23500    &      -13.2   \\
HD92850  &   B0Ia  &         28300    &             -13.5     &    BD+6073  &   B1Ib  &         24700    &         -14.0       \\
HD122879  &   B0Ia  &         27400    &            -13.1     &    BD+6476  &   B1Ib  &         23300    &         -13.6        \\
BD+631964  &   B0Iab  &         26500    &         -12.9    &    BD+6389  &   B1Ib  &         25400    &         -13.8    \\
CPD-573507  &   B0Iab  &         26800    &       -13.0      &    HD13659  &   B1Ib  &         24300    &         -13.6        \\
HD156212  &   B0Iab  &         28600    &         -13.4     &    HD15571  &   B1Ib  &         24700    &         -13.5   \\
HD167311  &   B0Ib  &         24700    &         -13.4       &     HD24398  &   B1Ib  &         23500    &         -13.7   \\
HD192660  &   B0Ib  &         27000    &         -13.0       &     HD47240  &   B1Ib  &         21900    &         -13.0     \\
HD205196  &   B0Ib  &         27100    &         -13.2      &   HD99857  &   B1Ib  &         24400    &         -14.0   \\
BD+6370    &   B0Ib  &         27100    &         -13.3      &       HD121228  &   B1Ib  &         23500    &         -14.0   \\
HD156134  &   B0Ib  &         28400    &         -13.3      &      HD119608  &   B1Ib  &         22000    &         -13.3      \\
HD164402  &   B0Ib  &         28900    &         -13.6     &      HD157246  &   B1Ib  &         23700    &         -14.0      \\
HD215806  &   B0Ib  &         32400    &         -13.2     &         HD158243  &   B1Ib  &         21200    &         -13.4   \\
HD171012  &   B0.2Ia  &         25900    &         -12.7    &   HD154385  &   B1Ib  &         26000    &         -13.8   \\
HD204172  &   B0.2Iab  &         27300    &         -13.3  &      HD155450  &   B1Ib  &         23600    &         -14.0   \\
HD16808  &   B0.2Ib  &         26300       &         -13.2  &    HD168750  &   B1Ib  &         25600    &         -13.9      \\
BD-124970  &   B0.5Ia  &         24600    &         -12.7  &      HD194279  &   B2Ia  &         19400    &         -13.0   \\
HD166787  &   B0.5Ia  &         25500    &         -13.6    &    HD14143  &   B2Ia  &         19900    &         -13.1    \\
HD169754  &   B0.5Ia  &         26500    &         -12.8     &     HD14818  &   B2Ia  &         20300    &         -13.2    \\
HD173987  &   B0.5Ia  &         26200    &         -13.2      &    HD41117  &   B2Ia  &         19800    &         -13.0      \\
HD194839  &   B0.5Ia  &         25800    &         -12.9  &      HD60308  &   B2Ia  &         18800    &         -13.3    \\
HD228712  &   B0.5Ia  &         25200    &         -13.0    &   HD62844  &   B2Ia  &         19000    &         -13.3  \\
HD219287  &   B0.5Ia  &         27000    &         -12.6    &    HD99953  &   B2Ia  &         20300    &         -13.1   \\
HD38771  &   B0.5Ia  &         26700    &         -13.3     &    HD134959  &   B2Ia  &         18800    &         -13.2   \\
HD152235  &   B0.5Ia  &         24000    &         -13.0    &     HD169034  &   B2Ia  &         15900    &         -13.3 \\
BD+60493  &   B0.5Ia  &         26100    &         -12.8   &    HD148379  &   B2Iab  &         18500    &         -13.0   \\
HD52382  &   B0.5Ia  &         23600    &         -13.2     &    HD154090  &   B2Iab  &         23400    &         -13.0     \\
HD152234  &   B0.5Ia  &         26100    &         -13.0    &  HD206165  &   B2Ib  &         19800    &         -13.3    \\
HD101545B  &   B0.5Iab  &         26900    &      -13.3   &  HD236740  &   B2Ib  &         18600    &         -14.0  \\
HD171432  &   B0.5Ib  &         25100    &         -13.6    &   HD41398  &   B2Ib  &         22000    &         -13.0    \\
HD192422  &   B0.5Ib  &         26300    &         -13.2    &   HD117024  &   B2Ib  &         22000    &         -14.0   \\
HD155985  &   B0.5Ib  &         25100    &         -13.6    &   HD116084  &   B2Ib  &         19400    &         -13.4    \\
HD64760  &   B0.5Ib  &         25000    &         -13.1     &   HD165024  &   B2Ib  &         22500    &         -14.0 \\
HD213087  &   B0.5Ib  &         25600    &         -13.3  &   HD172510  &   B2Ib  &         21900    &         -13.9   \\
HD344776  &   B0.5Ib  &         28000    &         -13.5   &  HD224055  &   B3Ia  &         16600    &         -13.3   \\
BD+612509  &   B0.5Ib  &         28300    &         -13.5   &  HD4694  &   B3Ia  &         16500    &         -13.9    \\
HD216411  &   B0.7Ia  &         23600    &         -12.7    &    HD14134  &   B3Ia  &         16700    &         -13.3  \\
HD2905  &   B0.7Ia  &         24400    &         -12.9     &   HD53138  &   B3Ia  &         16800    &         -13.4   \\
BD+353955  &   B0.7Iab  &         25000    &         -13.4   &  HD178129  &   B3Ia  &         16400    &         -14.0  \\
HD190066  &   B0.7Ib  &         24800    &         -13.3   &       HD240256  &   B3Ib  &         17100    &         -14.0 \\
HD91943  &   B0.7Ib  &         26700    &         -13.0     &     HD7103  &   B3Ib  &         16800    &         -13.5      \\
BD-175190  &   B1Ia  &         24500    &         -13.5     &    HD250290  &   B3Ib  &         15300    &         -14.0       \\
HD170938  &   B1Ia  &         23300    &         -13.0    &   HD51309  &   B3Ib  &         16800    &         -14.0    \\
HD58510  &   B1Ia  &         23600    &         -13.2     &  HD103338  &   B3Ib  &         17500    &         -14.0    \\
HD186841  &   B1Ia  &         26000    &         -13.2   &  HD43384  &   B3Iab  &         15900    &         -13.4       \\
HD142758  &   B1Ia  &         21600    &         -13.2   &           &          &                  &                \\
\end{longtable}
\FloatBarrier 
\clearpage

\end{appendix}

\end{document}